\documentclass{aa}

\usepackage{graphicx}
\usepackage{txfonts}
\usepackage{placeins}
\usepackage{subcaption}
\usepackage{subfiles}
\usepackage{xcolor}
\usepackage{lscape}
\usepackage{placeins}
\usepackage{multirow} 

\usepackage[colorlinks=true, citecolor=blue]{hyperref}

\usepackage[capitalize]{cleveref} 

\usepackage{xspace}
\newcommand{\fOneScore}{$F_1$\xspace}

\begin{document}

   \title{New substellar candidates identified through deep learning in the F150 sample of the large-scale SHINE direct imaging survey}

   \titlerunning{NA-SODINN candidates in SHINE}

   \author{C. Cantero\inst{1},
          M. Sabalbal\inst{2}, O. Absil\inst{2}\fnmsep\thanks{F.R.S.-FNRS Research Director}, M. Van Droogenbroeck\inst{3},  D. Ségransan\inst{1}, P. Delorme\inst{4}}

    \authorrunning{Cantero et al.}

    \institute{Geneva Observatory, University of Geneva, Chemin Pegasi 51, 1290 Versoix, Switzerland
    \and
    STAR Institute, Université de Liège, Allée du Six Août 19C, 4000 Liège, Belgium
    \and
    Montefiore Institute, Université de Liège, 4000 Liège, Belgium
    \and
    IPAG, Univ. Grenoble Alpes, CNRS, Grenoble, France}

   \date{Received September 15, 1996; accepted March 16, 1997}

  \abstract
   {The SPHERE High-contrast Imaging survey for Exoplanets (SHINE) represents one of the largest and most sensitive direct imaging campaigns conducted to date, targeting over 400 young, nearby stars with the goal of detecting and characterizing giant exoplanets and brown dwarfs. This extensive dataset offers a unique opportunity to revisit observations using modern, data-driven approaches, potentially uncovering new substellar candidates that may have been overlooked by classical analysis techniques.}
   {In this context, our study focuses on reprocessing and reanalyzing the so-called F150 sample, a well-defined subset of 150 main-sequence stars within 100 pc observed in the H-band with VLT/SPHERE as part of the SHINE survey.}
   {We apply NA-SODINN, a supervised deep learning model specifically tailored for detecting faint planetary signals in angular differential imaging (ADI) sequences. Designed to model local noise properties and capture spatial context, NA-SODINN is particularly effective at distinguishing real companions from residual speckle noise. To translate the model’s pixel-wise confidence maps into actionable detections, we introduce a novel F1-score-based thresholding strategy. This principled approach balances sensitivity and specificity, addressing a key limitation in current deep learning-based methods.}
   {NA-SODINN recovers all known companions and some of the debris disks in the F150 sample, and identifies 13 new substellar candidates not reported in previous studies: ten detected in both the H2 and H3 bands, and three in only one band. For the ten sources detected in both bands, we use the H2–H3 color–magnitude diagram to perform a first assessment of their nature. Based on this analysis, we identify two ambiguous cases and three photometrically promising candidates. However, in light of the currently available multi-epoch SPHERE data, only the candidate around Smethells 20 remains a strong target for follow-up.}
   {}


   \maketitle

    \section{Introduction}
    \label{sec:introduction}

Exoplanet direct imaging has become crucial for complementing indirect detection methods. Since the first direct detection of an exoplanet~\citep{Chauvin_2004}, the field of high contrast imaging (HCI) has become an essential tool not only for detecting young, self-luminous giant planets at wide separations (typically between 10 and
100 au) but also for enabling follow-up low-resolution spectroscopy of these objects, providing crucial insights into their chemical composition and atmospheric properties. Having access to this wide orbit space enables statistical analyzes that help constrain occurrence rates, shedding light on the initial conditions of protoplanetary disks~\citep{Janson_2021}, and serving as critical tracers of formation mechanisms and dynamical histories, ultimately refining current formation models \citep[e.g.,][]{Bowler_2016}. 

To date, HCI has been able to detect around 60 substellar companions around young nearby stars, mostly thanks to the implementation of HCI surveys over the past two decades. Early surveys, with sample sizes of 50–100 stars, led to the discovery of some of the first planetary-mass companions at large separations (>100 au) or with low mass ratios relative to their host stars \citep[i.e.,][]{Chauvin_2004, Chauvin_2005, Neuhaser_2005, lafreniere_2008}. These discoveries paved the way for the detection of the most iconic directly imaged planetary systems, including HR~8799~bcde~\citep{Marois_2008_HR8799} and $\beta$-Pictoris b~\citep{Lagrange2009AProbable}. More recent large-scale surveys, such as the Gemini Planet Imager Exoplanet Survey \citep[GPIES,][]{Nielsen2019TheGemini} and the SpHere INfrared survey for Exoplanets \citep[SHINE,][]{Chauvin_2017_shine}, have expanded their sample sizes to 600 and 400 stars, respectively. So far, these new generation surveys have only led to the discovery of five giants, 51 Eri b~\citep{Macintosh_2015}, HIP 65426 b~\citep{Chauvin_2017_HIP65426}, PDS 70 b and c~\citep{Keppler2018Discovery}, and HD143811 (AB)b~\citep{Squicciarini_2025}, and some higher mass brown dwarfs~\citep{Konopacky_2016, Cheetham_2018}. This relatively low number of HCI detections compared to expectations from radial velocity extrapolations \citep[i.e.,][]{Cumming_2008} is not solely indicative of a true absence of wide-orbit planets. Rather, it also highlights the substantial observational challenges, such as image contrast, that continue to limit the sensitivity of HCI surveys.

In HCI, contrast refers to the extreme flux ratio between a bright host star and its much fainter planetary companion. In the near-infrared,  typical contrasts for young giant planets range from $10^{-4}$ to $10^{-6}$. Achieving such contrasts using 8–10 meter ground-based telescopes requires a combination of precise instrumentation and advanced data processing. On the instrumental side, extreme adaptive optics (AO) systems and HCI instruments work in synergy during observations to correct for atmospheric turbulence and employ dedicated coronagraphs to suppress stellar light, respectively~\citep{Kenworthy_2025}. However, starlight is never perfectly removed in coronagraphic images. Residual wavefront errors and optical imperfections produce quasi-static speckles that can mimic planetary signals in both shape and brightness~\citep{Males_2021}. These limitations naturally lead to the algorithmic side, where observational strategies combined with image post-processing techniques are used to suppress these residual speckles, reach deeper contrasts, and recover companion signatures. The most widely used observing strategy on ground-based telescopes is angular differential imaging \citep[ADI,][]{Marois2006Angular}, in which coronagraphic images are acquired in pupil-tracking mode: the telescope pupil remains fixed while the image
rotates over time due to the Earth’s rotation. As a result, speckles associated with the telescope and instrument optical train remain mostly fixed in the focal plane while the astrophysical signals rotate around the star as a function of the parallactic angle. This generates a sequence of coronagraphic images (ADI sequence, hereafter) that contains rich angular diversity and allows post-processing algorithms to effectively disentangle true planetary signals from residual image noise. 

Over the past two decades, numerous image post-processing techniques have been developed to improve planet detection in ADI datasets \citep[see][for a review]{Cantalloube_2020}. Among these, a major family relies on point spread function (PSF) subtraction. These methods aim to build a model of the stellar PSF from the ADI sequence and subtract it from each frame. The residuals are then derotated and combined based on their parallactic angles, producing a final image in which residual speckles are further attenuated and spatially decorrelated. This output is referred to as the processed frame. Among PSF subtraction techniques, principal component analysis \citep[PCA,][]{Soummer2012Detection, Amara2012Pynpoint} and its extension, annular PCA~\citep{Absil2013Searching, Gomez2016LowRank}, have become standard approaches in major HCI surveys \citep[e.g.,][]{Nielsen2019TheGemini, Janson_2021, Langlois_2021_shine}. Their widespread use is largely due to several practical advantages: PCA is computationally efficient, has a solid mathematical foundation, requires minimal hyperparameter tuning~\citep{Bonse_2024}, and is currently integrated into major HCI pipelines such as VIP~\citep{Gomez_2017, Christiaens2023VIP}, PynPoint~\citep{Stolker_2019}, and pyKLIP~\citep{Wang_2015}. Recently, machine learning, and in particular deep learning, has emerged as a powerful alternative that can be adapted to work in synergy with PSF subtraction techniques like PCA, delivering superior performance in tasks such as exoplanet signal detection \citep[e.g.,][]{Gomez_2018_sodinn, Flasseur_2024_deepPACO} and speckle noise modeling~\citep{Yip2020Pushing, Wolf_2024}. A notable example is NA-SODINN~\citep{Cantero_2023}, a recent supervised binary classifier based on convolutional neural networks (CNNs). It enhances detection sensitivity on ADI-PCA processed frames by learning deep spatio-temporal correlations of residual speckle noise across different noise regimes and incorporates signal-to-noise (S/N) curves as additional predictors, which provide complementary information about companion-like features. This multi-dimensional learning strategy has enabled NA-SODINN to outperform traditional techniques, securing top ranks in the first phase of the Exoplanet Imaging Data Challenge~\citep[EIDC,][]{Cantalloube_2020}. 

In this study, we aim to leverage the power of NA-SODINN to reprocess the F150 sample from the SHINE survey in search of new substellar candidates. To improve its applicability to large-scale survey analysis, we introduce two key modifications to the original pipeline: (1) an automated, \fOneScore-score-driven method for selecting optimal detection thresholds in each output map, enhancing interpretability and improving candidate selection; and (2) contrast curve generation to quantify detection sensitivity across the survey. Aside from these additions, the core algorithm remains unchanged.  Thus, this work represents the first application of NA-SODINN to real data from a large-scale HCI survey, providing not only an opportunity to search for new companions but also to critically assess the algorithm’s performance, limitations, and future development paths.

This paper is structured as follows: \cref{sec:SHINE} introduces the SHINE survey and its observing strategy, with a particular focus on the F150 sample used in this study. It also summarizes the current status of the full survey and the complementary snapSHINE program. \Cref{sec:data_reduction} details the full data reduction process of the F150 sample, including image preprocessing, an overview of the NA-SODINN algorithm, the new thresholding strategy for candidate identification, the generation of final detection maps for each target, and the sensitivity analysis through contrast curves. \Cref{sec:results} presents the detected candidate companions and their analysis. Finally, \cref{sec:conclusions} discusses the implications of our results and summarizes the main conclusions.

    \section{SHINE survey}
    \label{sec:SHINE}

\begin{figure*}
    \centering
    \includegraphics[width=0.98\linewidth]{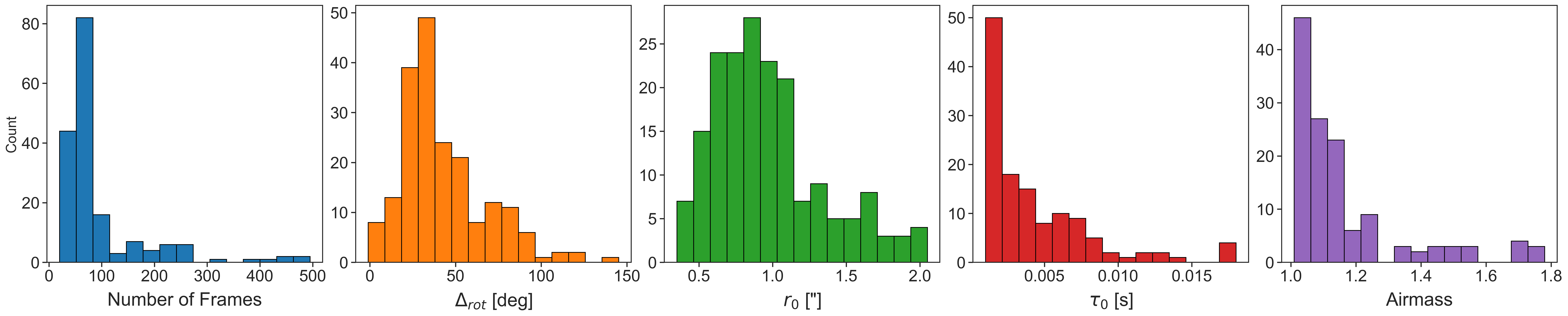}
    \caption{Distributions of the main features extracted from the ADI sequences of the F150 sample. The number of frames in the sequence (blue) corresponds to the final value obtained after applying the Pearson correlation analysis described in \cref{sec:pre-processing}. $\Delta_{\mathrm{rot}}$ (orange) refers to the total field rotation during the sequence. The three rightmost panels correspond to atmospheric conditions: $r_{0}$ (green) indicates the mean seeing, $\tau_{0}$ (red) the mean coherence time, and airmass (purple).}
    \label{fig:F150_histograms}
\end{figure*}

The SHINE survey~\citep{Chauvin_2017_shine} represents one of the most ambitious large-scale direct imaging campaigns. Conducted over 200 nights between February 2015 and September 2021, the survey was carried out in visitor mode at the Very Large Telescope (VLT) using the Spectro-Polarimeter High-contrast Exoplanet REsearch \citep[SPHERE,][]{Beuzit2019SPHERE}, combining its SAXO extreme adaptive optics
system~\citep{Fusco2006High-order} and its apodized pupil Lyot coronagraph \citep[APLC,][]{Soummer2005Lyot}. It targeted a statistically robust sample of over 400 young, nearby stars, with a comprehensive set of scientific objectives. These included the detection of new giant exoplanets and circumstellar disks, the characterization of the atmospheres of both newly discovered and previously known exoplanets, the investigation of giant planet formation mechanisms, and the study of the dynamical architecture of planetary systems.

Each observation spanned approximately one to two hours employing ADI. Observations were conducted in one of two SPHERE configurations: IRDIFS or IRDIFS-EXT. Both modes enable the simultaneous use of the InfraRed Dual-beam Imager and Spectrograph \citep[IRDIS,][]{Dohlen_2008_IRDIS} and the Integral Field Spectrograph \citep[IFS,][]{Claudi_2008, Mesa_2015_IFS}, offering complementary spectral coverage across near-infrared wavelengths. In IRDIFS mode, IFS captures data in the YJ spectral bands within a compact $1\farcs77 \times 1\farcs77$ field of view, while IRDIS operates in Dual-Band Imaging mode \citep[DBI,][]{Vigan_2010} using the H2-H3 filter pair (DB-H23), providing a broader $11\arcsec \times 11\arcsec$ field of view. In the extended IRDIFS-EXT mode, IFS expands its coverage to include the YJH bands, and IRDIS switches to the K1-K2 filter pair (DB-K12). This latter setup is particularly suited for detecting red, young L-type planets in nearby star-forming regions, such as HD~95086~b in Sco-Cen~\citep{Chauvin_2018}.

To date, SHINE has led to various discoveries: the giant exoplanet HIP 65426 b~\citep{Chauvin_2017_HIP65426}, the first unambiguous detection of the protoplanet PDS 70 b embedded in a protoplanetary disk ~\citep{Keppler2018Discovery}, one brown dwarf \citep[HIP 64892 B,][]{Cheetham_2018}, four new disk detections~\citep{Lagrange_2016_discovery_disk, Feldt_2017_discovery, Sissa_2018_discovery, Perrot_2019_discovery}, as well as 78 new stellar companions~\citep{Bonavita_2022_discovery}.

\subsection{The F150 sample\label{sec:F150_sample}}
An initial statistical analysis was performed on the first 150 targets observed between 2015 and 2017, referred to as the F150 sample\footnote{The table with all the ADI sequences and their main characteristics, including atmospheric conditions, is available at the CDS via \url{https://cdsarc.cds.unistra.fr/viz-bin/cat/J/A+A/706/A275}}.   \cref{fig:F150_histograms} shows the distribution of the main characteristics of the F150 sample. This subset was selected to be representative of the full survey and formed the basis of a trilogy of papers published in 2021.

In the first paper, \citet{Desidera_2021_shine} analyzed in detail the stellar properties of the F150 sample by combining indicators such as kinematics, moving group membership, lithium absorption, rotation periods, and chromospheric activity to estimate stellar ages and masses. The sample spans a broad range of spectral types, from B to M stars, with 53 BA stars, 77 FGK stars, and 20 M-type stars. The median age is 45 Myr, with 90\% of the sample falling between 11 and 450 Myr. Stellar masses range from 0.57 to 2.37 $M_\odot$, with a median of 1.15 $M_\odot$, and the median distance is approximately 48 pc. 

The observational setup and data processing for the F150 sample were presented in a second paper by \citet{Langlois_2021_shine}. In total, 16 substellar companions were identified in the sample, eight brown dwarfs and eight planetary-mass companions, with most having been previously reported by earlier direct imaging campaigns. Additionally, 1483 candidate point sources were detected in the IRDIS field of view across all epochs using PCA through the SpeCal pipeline. Of these, 1176 were confirmed as background sources through follow-up observations or archival data, while 307 remained single-epoch detections and were considered as new candidates. Notably, more than 95\% of these detections were obtained with the H2-H3 filter pair (DB-H23). Candidate prioritization was based on color-magnitude diagrams (CMDs), which led to the rejection of 44\% of the H23 candidates as likely contaminants; ambiguous cases were scheduled for second-epoch follow-up.

Building upon these detection results, \citet{Vigan_2021_shine} carried out in a third paper a statistical analysis aimed at constraining the occurrence rates of wide-separation giant planets. By converting contrast curves to mass limits using evolutionary models and comparing the results with both parametric and population synthesis models, they derived occurrence rates across different spectral types.

\subsection{Current status\label{sec:current_status}}
With the completion of the initial F150 sample analysis, the SHINE survey advanced to its full statistical phase, expanding the sample to 400 stars and accumulating over 700 high-contrast datasets. This enlarged dataset significantly increased the number of detected point sources, mostly single-epoch detections with IRDIS in the H2-H3 filter pair. To discriminate between physical companions and background contaminants, the snapSHINE program was proposed. Unlike the deeper SHINE exposures designed for discovery and spectral characterization, snapSHINE was conceived as a rapid second-epoch astrometric follow-up survey. It used shorter exposure times and simplified observing sequences, enabling the confirmation or rejection of large numbers of candidates with minimal telescope time. While less sensitive than SHINE, snapSHINE was optimized to deliver reliable astrometric measurements at separations beyond $\sim0$.5", where most single-epoch sources are located. 

As part of a new series of SHINE publications analyzing the full dataset, \citet{Chomez_2025_shine} recently presented the observing strategy, data quality assessment, and point source analysis for the combined SHINE and snapSHINE samples. In a similar spirit to the earlier work of \citet{Langlois_2021_shine} on the F150 data, this new analysis applies the PACO-ASDI post-processing algorithm, which enhances contrast limits by up to two magnitudes. More than 3500 point sources were detected and classified in this (re)processing effort. Although no new confirmed exoplanets were identified, the improved sensitivity revealed 24 additional promising sources. 

More recently, \citet{Sabalbal_2025} reprocessed the F150 sample using RSM~\citep{Dahlqvist2020Regime}, a post-processing algorithm that achieved similarly top-ranked detection performance to NA-SODINN in the EIDC~\citep{Cantalloube_2020}. They reported a gain in median sensitivity of about a factor of two at $\sim1$\arcsec\ and of $\sim4$–5 at smaller angular separations relative to standard PCA, reaching performance comparable to PACO on the same datasets. Additionally, RSM yielded 87 detected signals, including 38 not recovered in the PACO-based analysis, and highlighted one particularly promising single-epoch candidate.
    \section{Methodology}
    \label{sec:data_reduction}

This section describes the complete data reduction process applied to each ADI sequence of the F150 sample. It begins with the initial SPHERE pre-processing steps, continues through the generation of final detection maps using the NA-SODINN algorithm, and concludes with the assessment of the survey’s sensitivity based on contrast curve analysis.

\subsection{Raw data pre-processing}
\label{sec:pre-processing}
The SHINE F150 sample was pre-processed with the SPHERE data reduction and handling software~\citep{Pavlov_2008}, implemented at the High-Contrast Data Centre\footnote{\url{https://hc-dc.cnrs.fr/?rubrique16&lang=en}} \citep[HC-DC,][]{Maire_2016,Delorme_2017, Galicher_2018, Beuzit2019SPHERE}. The initial reduction steps for the IRDIS data involved standard calibrations, including dark subtraction, flat-fielding, bad pixel correction, and background subtraction. Precise centering of the star was achieved using four satellite spots imprinted on the IRDIS images via a specific pattern applied to the deformable mirror. For each target in the sample, the output of this pre-processing is the ADI sequence with re-centered frames (individual detector integrations), along with the corresponding list of de-rotation angles required for ADI post-processing. For more details, including SPHERE calibration, we refer to \citet{Langlois_2021_shine}.

\subsection{Image processing}
\label{sec:image-proc}
For this study, we considered two additional operations on each ADI sequence. First, each frame in the sequence is cropped to a 190 $\times$ 190 pixel region centered on the star, corresponding to a field of view radius of approximately $1\farcs2$. This inner region aligns with the area most effectively corrected by the SAXO extreme-AO system and where NA-SODINN has shown superior performance compared to standard PSF subtraction techniques~\citep{Cantero_2023}. Second, using the \texttt{VIP} Python package~\citep{Gomez_2017, Christiaens2023VIP}, low-quality frames are identified and removed by computing the Pearson correlation between each frame and the median frame in the sequence. Frames with correlation values below 0.8 are discarded, as they potentially correspond to frames affected by poor AO correction or fast atmospheric variations. This threshold was chosen as a good trade-off between preserving data volume and filtering out frames that degrade post-processing performance. Both operations contribute to reducing the computational cost for the post-processing.

\begin{figure*}[ht!]
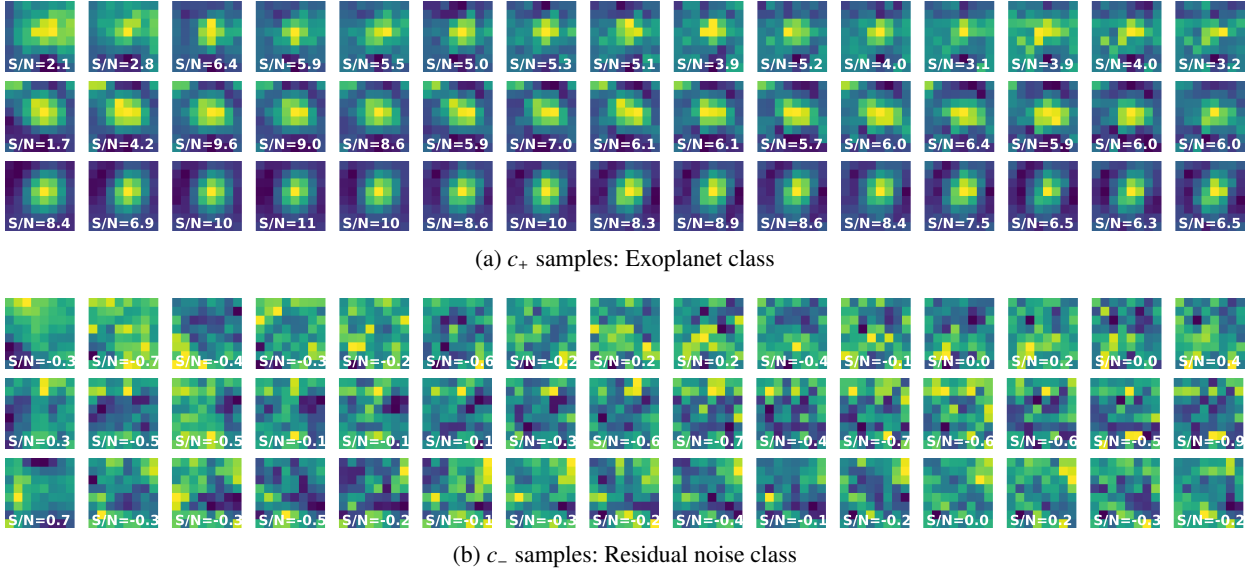

    \centering
    \begin{subfigure}{0.89\textwidth}
        \centering
        \includegraphics[width=\textwidth]{figures/cplus_low.png}\vspace{1mm}
        \includegraphics[width=\textwidth]{figures/cplus_medium.png}\vspace{1mm}
        \includegraphics[width=\textwidth]{figures/cplus_high.png}
        \caption{$c_{+}$ samples: Exoplanet class}
        \label{fig:cplus}
    \end{subfigure}

    \vspace{3mm} %

    \begin{subfigure}{0.89\textwidth}
        \centering
        \includegraphics[width=\textwidth]{figures/cminus1.png}\vspace{1mm}
        \includegraphics[width=\textwidth]{figures/cminus2.png}\vspace{1mm}
        \includegraphics[width=\textwidth]{figures/cminus3.png}
        \caption{$c_{-}$ samples: Residual noise class}
        \label{fig:cminus}
    \end{subfigure}
    
    \caption{Illustration of the training set used by NA-SODINN, showing three examples of patch sequences from the $c_{+}$ class (exoplanet, top) and three from the $c_{-}$ class (residual noise, bottom). Each sample consists of a sequence of $N_k = 15$ image patches, each extracted from an ADI-PCA processed frame obtained with a different number of principal components, from $k=2$ to $k=30$ in steps of two. The S/N value displayed inside each patch is the S/N measured at the corresponding $k$, and the collection of these values across the sequence forms the associated S/N curve. For the $c_{+}$ class, the three samples are ordered by increasing injected flux, with injections selected such that the initial S/N at the first principal component $k = 1$ lies in the range 2--4. As $k$ increases, the S/N typically rises at first as PCA better models and removes speckle noise, and then may decrease at higher $k$ when the algorithm starts to subtract the planetary signal partially (self-subtraction). The patch size is set to twice the PSF FWHM (4 pixels in this case), resulting in $8\times 8$ pixel patches.}
    \label{fig:training_samples_example}
\end{figure*}

\subsection{Advanced ADI post-processing}
\label{sec:ovierview}

To search for companion candidates in each ADI sequence, we use the NA-SODINN~\citep{Cantero_2023} post-processing algorithm, standing for Noise-Adaptive Supervised exOplanet detection via Direct Imaging with a deep Neural Network. It represents an upgraded version of SODINN~\citep{Gomez_2018_sodinn}, its predecessor, and is a binary classifier model designed to distinguish between companion point-like signatures and residual noise in annular PCA residual frames (ADI-PCA processed frames, hereafter). In this section, we provide a brief overview of the NA-SODINN processing pipeline, which consists of four main steps. For a more detailed description, we refer to \citet{Cantero_2023}.

\subsubsection{First step: Identification of noise regimes}
\label{sec:noise_regimes}

Given an ADI sequence, the computation starts with a preliminary analysis of the residual noise distribution across the ADI-PCA processed frame. While this residual noise has often been (implicitly) assumed to be Gaussian ---an assumption that produces high false positive detection rates~\citep{Marois2008Confidence, Mawet2014Fundamental} since it is not valid at small angular separations~\citep{Pairet2019STIM, Dahlqvist2020Regime, Daglayan_2022, Daglayan_2024}--- NA-SODINN does not explicitly assume a specific probability distribution. Instead, it acknowledges that independent of the exact noise distribution, two non-overlapping noise regimes generally exist with different statistical properties: a non-Gaussian regime at small separations, dominated by residual speckle noise, and a Gaussian regime at larger separations, where speckle noise becomes smaller than background noise. NA-SODINN identifies and spatially delimits these noise regimes within the ADI-PCA processed frame. As detailed later in  \cref{sec:training_set}, this stratification strategy of the field of view enables the generation of independent training sets that accurately capture the statistical correlations of each noise distribution. 

\subsubsection{Second step: Generation of training sets}
\label{sec:training_set}

Once the noise regimes are identified,  NA-SODINN proceeds with the generation of the training sets. For each regime, an independent dataset is created, consisting of two classes: the exoplanet class ($c_{+}$) and the noise class ($c_{-}$). Both classes are represented by hundreds of thousands of sequences of square image patches along with their corresponding signal-to-noise (S/N) curves (\cref{fig:training_samples_example}). These samples are generated annulus by annulus within each considered regime.

A $c_{+}$ patch sequence is generated through three consecutive actions. First, the instrumental PSF is injected\footnote{In HCI, a planetary injection is defined as the process of pasting the AO-corrected instrumental PSF (centered, cropped, and scaled to the desired contrast) to every frame in the ADI sequence at specific coordinates following field rotation.} at a random pixel within the considered noise regime. The flux of the injection is randomly selected such that the injection yields a S/N between 2 and 4 in the ADI-PCA processed frame produced with one single principal component. This modified ADI sequence is then processed using annular PCA with multiple $k$-subtraction levels (i.e., varying numbers of principal components). This results in $k$ different ADI-PCA processed frames. Finally, from each of these, a square patch is cropped around the injection coordinates, forming a sequence of $k$ patches that together constitute a single $c_{+}$ training sample (\cref{fig:cplus}). This sample contains local spatio-temporal pixel correlations of the synthetic injection that represent its evolution in PCA space (different $k$-subtraction levels). The associated S/N curve is computed following the definition from \citet{Mawet2014Fundamental}, measuring the S/N of the injection at each $k$-subtraction level. This curve encodes additional information about the interaction between the injected signal and the residual noise in the rest of the annulus~\citep{Cantero_2023}. 

A $c_{-}$ patch sequence and its S/N curve (\cref{fig:cminus}) are generated using the same procedure, except that no fake companion is injected. Since each ADI sequence yields only one realization of residual noise (per pixel), the number of $c_{-}$ samples is limited to the total number of pixels in the annulus, which increases the risk of overfitting. To deal with this issue, NA-SODINN makes use of data augmentation techniques such as rotations, averages, and shifts to produce $c_{-}$ training samples. As a result, these samples capture similar spatial and temporal correlations to the $c_{+}$ class, but purely due to residual noise. The same number and order of $k$-subtraction levels are used as for the $c_{+}$ class to ensure the same noise subtraction in both classes.

\subsubsection{Third step: Training}
\label{sec:model_training}
NA-SODINN continues by training a separate detection model for each noise regime using its corresponding training set. The network consists of two concatenated convolutional-LSTM blocks~\citep{Shi2015ConvLSTM} with spatial 3D dropout~\citep{Srivastava2014Dropout} and MaxPooling-3D layers~\citep{Boureau2010ATheoretical}, extracting spatio-temporal correlations from patch sequences. The resulting feature maps are flattened and concatenated with the corresponding S/N curves, forming a vector that also encodes S/N correlations. This vector is passed through a fully connected layer with 128 hidden units, a rectified linear unit (ReLU) activation function~\citep{Nair2010Rectified}, and dropout regularization. The output layer consists of a sigmoid unit, providing a confidence score between 0 and 1. The network weights are initialized using a Xavier uniform initializer and optimized via backpropagation with a binary cross-entropy loss function. An Adam optimizer is employed with a learning rate of 0.003 and mini-batches of 64 training samples. Training is guided by an early stopping condition based on validation loss, with typically 7–10 epochs yielding $\sim99.9$ validation accuracy.

\subsubsection{Fourth step: Confidence map}
\label{sec:inference}
\begin{figure*}
    \centering
    \includegraphics[width=0.3\linewidth]{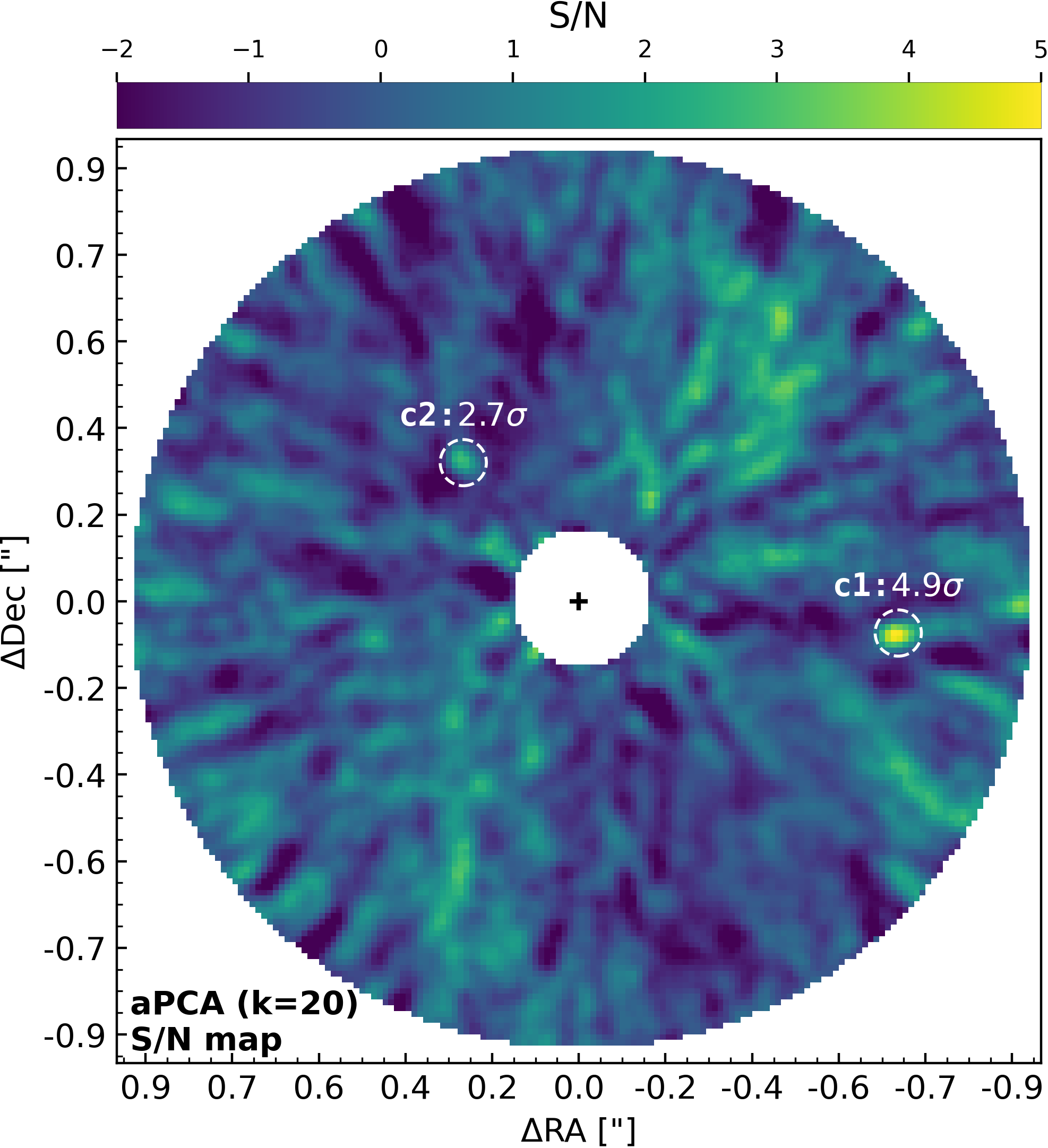}\hspace{0.5cm}
    \includegraphics[width=0.3\linewidth]{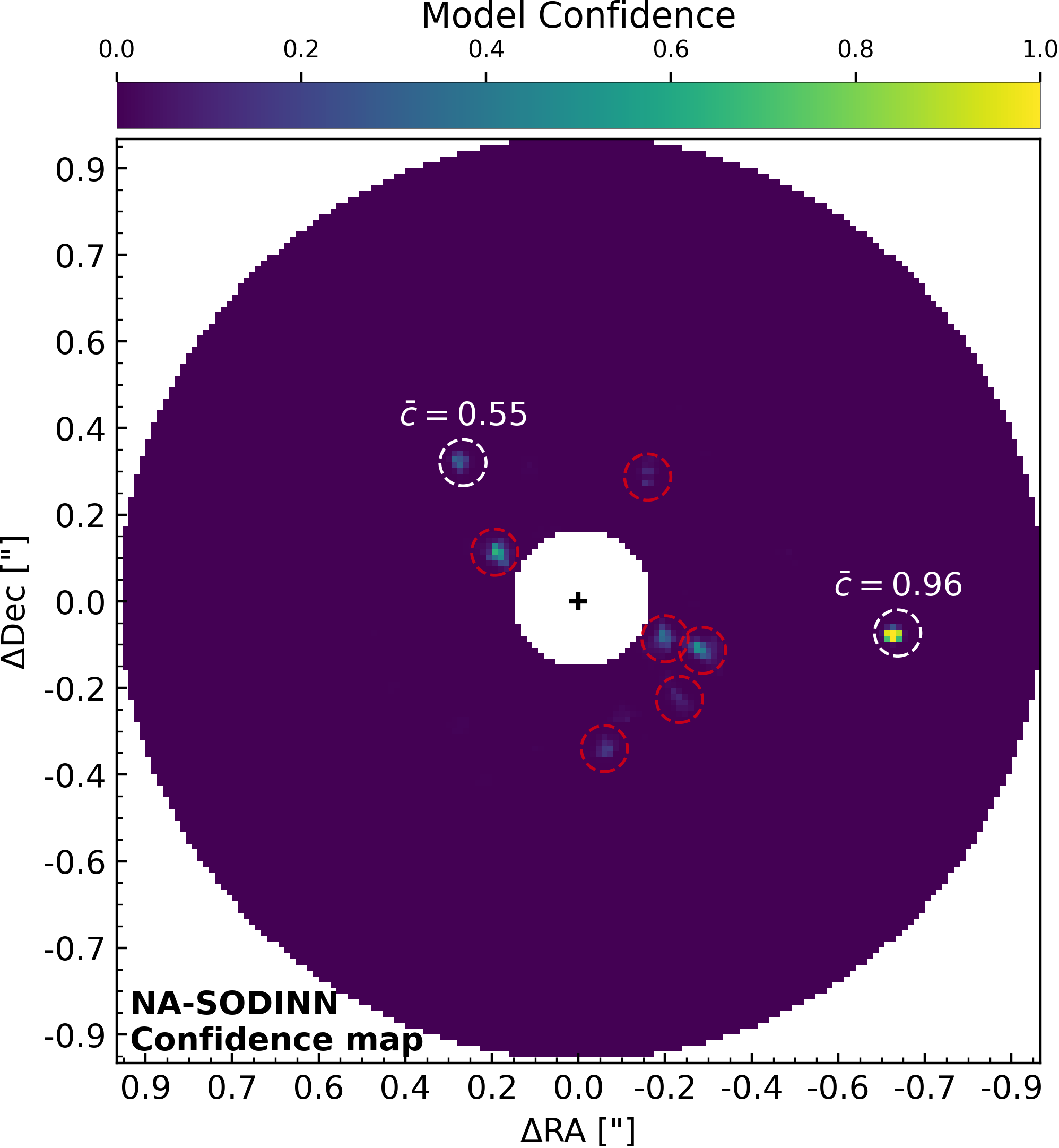}
    \caption{Comparison between the S/N map (left), computed using annular PCA with $k = 20$ principal components, and the confidence map (right), computed using NA-SODINN trained with faint companion injections, for the \textit{sph1} SPHERE ADI sequence from EIDC~\citep{Cantalloube_2020}. The small dashed circles in white indicate the true locations of the injected fake companions c1 and c2, while the red circles mark the positions of all NA-SODINN false positives if the detection threshold was set to a confidence score of 0.4 (see text). For both c1 and c2 injections, the S/N map highlights their sigma detection level while the confidence map shows the median confidence of all pixels located within a $1-\lambda/D$ aperture. The circular concentric masked area with a radius of $\sim3\lambda/D$ at the center of each image indicates the inner working angle.}
    \label{fig:confidence_map_comparison}
\end{figure*}

The final stage of NA-SODINN involves applying the trained models to detect real companions in the same ADI sequence used for training\footnote{In NA-SODINN, training and inference are performed on the same ADI sequence to ensure consistency of residual noise statistics. Since noise regimes are empirically defined and dataset-dependent, using external ADI sequences to build the $c_{-}$ class and increase noise diversity could introduce domain mismatch effects. Noise diversity is instead achieved through data augmentation of the $c_{-}$ class.}. In the same fashion as the training set generation, inference is performed annulus by annulus within each corresponding noise regime. Given a pixel coordinate within a regime's annulus, NA-SODINN first extracts the associated patch sequence and S/N curve. These inference samples are fed into the trained model corresponding to the pixel’s regime, which returns a confidence score between 0 and 1. This score reflects the model's confidence that the pixel belongs to the $c_{+}$ class in such a way that higher scores (closer to 1) indicate likely companion detections, and lower scores point to regions dominated by residual noise. Repeating this process across the full field of view produces the confidence map: a pixel-level image aligned with the target’s field of view (i.e., \cref{fig:confidence_map_comparison}-right), where companion detection can be performed by defining a confidence threshold.

\subsection{Interpreting confidence maps \label{sec:model_interpretation}}

\subsubsection{The confidence calibration problem}
NA-SODINN is a data-driven framework designed to detect companion signatures in ADI sequences. This approach aims to offer two main advantages with respect to more standard post-processing techniques: (1) it achieves an improved balance between sensitivity and specificity across all angular separations and noise regimes \citep[see][]{Cantero_2023}, and (2) it aims to enhance interpretability through confidence maps that, unlike traditional processed frames or their S/N maps, provide a more homogeneous and reproducible basis for identifying candidate companions across different angular separations and noise regimes. Although both approaches ultimately rely on thresholding, confidence maps are based on patterns learned directly from the data rather than on residual intensities or S/N values alone.

However, confidence maps must be interpreted with caution. The confidence scores produced by binary classifiers, while bounded between 0 and 1 via a sigmoid activation (\cref{sec:model_training}), do not represent calibrated probabilities~\citep{Tao_2023} and therefore cannot be interpreted as such. Instead, they reflect the model’s internal certainty, which can be misleading. As a result, selecting an appropriate detection threshold cannot be based solely on intuition or on the assumption that a high score equates to a high probability of a true companion. This limitation becomes more pronounced when we aim to process large HCI surveys, such as the F150 sample, where NA-SODINN may flag numerous candidate companions with varying confidence levels. 

\Cref{fig:confidence_map_comparison} illustrates this challenge, using the \textit{sph1} SPHERE ADI sequence from the Exoplanet Imaging Data Challenge~\citep{Cantalloube_2020}. The left panel shows the S/N map obtained via annular PCA, where the two injected companions are highlighted with white circles: c1, located at a separation of $\sim650$ mas with a significance of $4.9\sigma$, and c2, at $\sim300$ mas with a lower value of $2.7\sigma$. In the corresponding NA-SODINN confidence map (right panel), c1 receives a high model confidence score $\bar{c} \sim 0.96$, while c2 is assigned a moderate score $\bar{c} \sim 0.55$. At first glance, this behavior might suggest a simple monotonic relation between apparent signal strength and model confidence: brighter companions receive higher scores, and fainter ones receive lower scores. However, the confidence map also contains several additional bright spots at similar separations (marked with red dashed circles) that are not associated with any injected companion. These false positives receive confidence scores in the same 0.4–0.6 range as c2, making them indistinguishable from a real detection based solely on raw confidence. This overlap illustrates a key issue: intermediate confidence values do not correspond to calibrated probabilities and cannot be directly interpreted as such. Without a principled thresholding mechanism, users are left to subjectively decide whether a pixel value like 0.55 represents a true companion or residual noise.

\begin{figure*}
    \centering
    \includegraphics[width=0.3\linewidth]{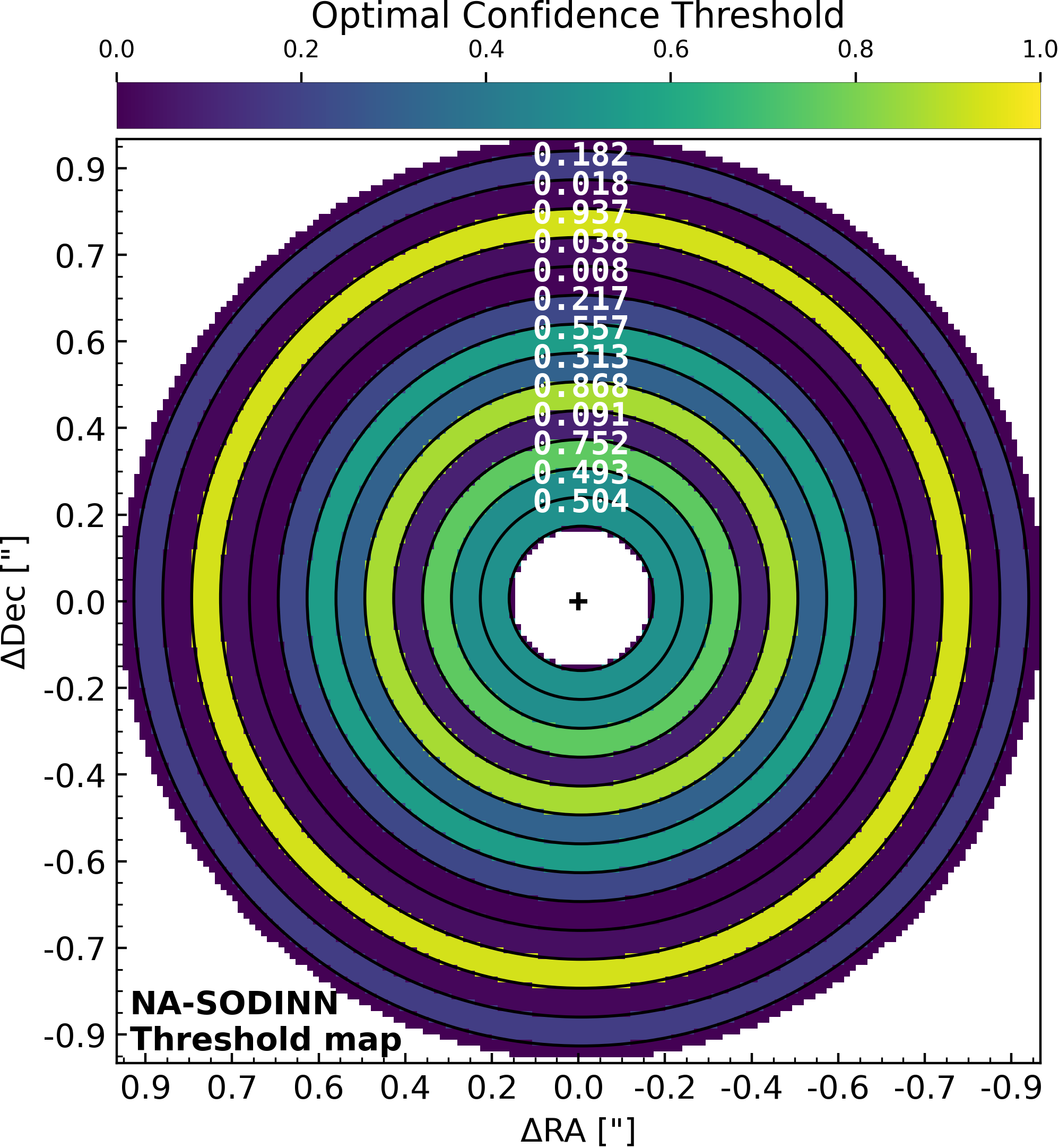}\hspace{0.5cm}
    \includegraphics[width=0.3\linewidth]{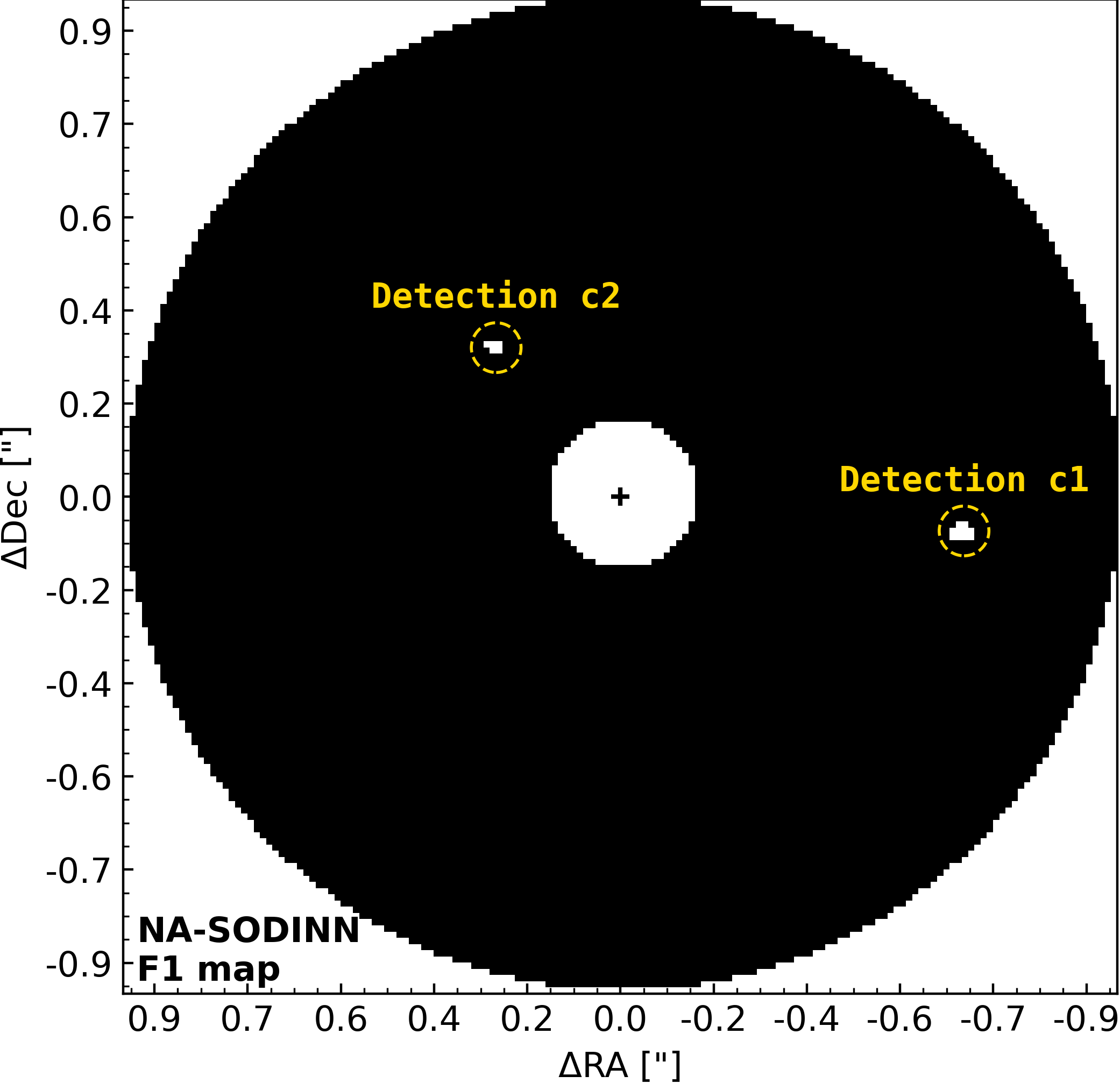}
    \caption{Output of NA-SODINN’s proposed thresholding strategy for the \textit{sph1} case shown in \cref{fig:confidence_map_comparison}. \textit{Left}: Threshold mask indicating the optimal confidence threshold per annulus, i.e., the value that maximizes the mean \fOneScore score across injection-recovery realizations. The overlaid white labels complement the colour map by showing the selected thresholds at each annulus. \textit{Right}: Final binary detection map obtained by applying the threshold mask to the original confidence map of \cref{fig:confidence_map_comparison}.}
    \label{fig:threshold_detection_map}
\end{figure*}

Several techniques have been proposed in the deep learning community to address this calibration problem. Overall, these can be categorized into three main groups. The first consists of post-hoc calibration methods, which aim to adjust model predictions after training by learning additional parameters on a held-out calibration dataset with the same properties as the training set. Examples include Platt scaling~\citep{Platt_1999}, isotonic regression~\citep{Zadrozny_2002}, histogram binning~\citep{Zadrozny_2001}, Bayesian binning~\citep{Naeini_2015}, beta calibration~\citep{Kull_2017}, and temperature scaling~\citep{Guo_2017}. The second category involves regularization-based approaches that aim to improve calibration indirectly by enhancing model generalization. These include data augmentation techniques like Mixup~\citep{Thulasidasan_2019} and AugMix~\citep{Hendrycks_2020}, ensemble methods~\citep{Lakshminarayanan_2017}, and explicit regularization strategies such as weight decay~\citep{Guo_2017}. Finally, the third category focuses on modifying the training loss function by introducing an auxiliary differentiable surrogate loss to approximate expected calibration error~\citep{Karandikar_2021}, or replacing the standard cross-entropy loss with alternatives such as mean squared error~\citep{Hui_2021}, inverse focal loss~\citep{Wang_2021}, or focal loss variants~\citep{gupta_2021}. 

While these approaches offer promising avenues for improving the interpretability of confidence maps, their adaptation to NA-SODINN is beyond the scope of this paper. Therefore, rather than interpreting confidence scores as probabilities, we adopt an alternative strategy tailored to the NA-SODINN architecture and for the analysis of the F150 sample: a threshold-optimization approach that seeks to identify, per annulus, the confidence value that maximizes detection performance. Specifically, this means selecting the confidence threshold that optimizes the trade-off between false positives (FP) and false negatives (FN), 
on a per-dataset basis.

\subsubsection{Defining robust confidence thresholds }
\label{sec:f1_threshold}

In its original form, NA-SODINN performs inference annulus by annulus, scoring all pixels in the ADI sequence using the trained model corresponding to the local noise regime (\cref{sec:inference}). Once all pixels in an annulus are scored, NA-SODINN immediately proceeds to score all pixels in the next annulus without evaluating the reliability of the resulting confidence scores. In the proposed approach, this gap is addressed by introducing a dedicated injection-recovery step. 

After scoring all pixels in a given annulus, NA-SODINN temporarily pauses inference and generates a set of $n = 50$ synthetic cubes, each created by duplicating the original ADI sequence and injecting two fake companions with random fluxes and positions restricted to the same annular region. Inference is then re-run on each of these $n$ synthetic cubes, computing confidence scores only for the pixels within the target annulus. To assess detection performance, a dense grid of confidence thresholds between 0 and 1 is defined. For each threshold in the grid, the model’s outputs are compared with the known injection positions, allowing for the calculation of true positives (TPs), false positives (FPs), and false negatives (FNs). These basic indicators are then used to compute the \fOneScore-score (see \cref{app:F1_score}) at each threshold. This procedure yields an \fOneScore-score curve for every $n$ realization, characterizing model performance as a function of the confidence threshold, as described in \citet{Sabalbal_2025}. Because the same number of fake companions is injected in each synthetic cube, the resulting individual \fOneScore scores are computed under the same number of positive priors, which allows their averaging to remain meaningful (\citealt{Pierard2020Summarizing}). Averaging the $n$ individual \fOneScore curves therefore provides a robust estimate of detection performance across thresholds, and the confidence value at which the mean curve reaches its maximum is selected as the optimal threshold for that annulus, as it offers the best trade-off between detection completeness and reliability.

This injection-recovery process is repeated independently for each annulus across the field of view. For each annulus, the confidence threshold corresponding to the maximum of the mean \fOneScore curve is selected as the optimal value. These optimal thresholds are then used by NA-SODINN to construct a threshold map, an annulus-wise mask that assigns to each pixel the confidence threshold associated with its angular separation (\cref{fig:threshold_detection_map}, left). Applying this spatially varying threshold to the NA-SODINN confidence map yields what we refer to as the \fOneScore map: a binary detection map in which activated pixel blobs represent regions where the underlying signal meets the model’s strongest internal criteria for being distinguishable from noise, given the local noise regime and learned data characteristics. This approach resolves the ambiguity inherent in raw, uncalibrated confidence values by providing a clear and objective decision rule for detection. \Cref{fig:threshold_detection_map}-right shows the resulting \fOneScore map for the \textit{sph1} dataset, which successfully recovers both injected companions (c1 and c2) without introducing any false positives.

\subsection{Contrast curves\label{sec:contrast_curves}}
To assess the detection limits achieved by a post-processing algorithm in an HCI survey, contrast curves are typically computed from S/N maps using the standard $5\sigma$ criterion \citep[e.g.,][]{Mawet2014Fundamental, Langlois_2021_shine, Chomez_2025_shine}. In this approach, contrast is estimated at each angular separation from the computation of the noise standard deviation and the throughput. 

NA-SODINN, however, produces confidence maps that assign a confidence score to each pixel rather than a flux-based S/N ratio. Since these confidence values do not follow the statistical assumptions underlying the classical $5\sigma$ framework, the standard definition of contrast cannot be applied consistently. Instead, we adopt an empirical completeness-based definition, following the injection-recovery philosophy of \citet{Dahlqvist2021Improving}. For each annulus, we first fix the detection threshold to the annulus-dependent confidence threshold derived in \cref{sec:f1_threshold}, which is the value that maximizes the  \fOneScore-score. At each angular separation, 20 synthetic companions are then injected at different azimuths, and their fluxes are iteratively adjusted until NA-SODINN recovers 95\% of them at this fixed threshold. A synthetic companion is considered recovered if it is detected at the expected location in the resulting binary detection map. The corresponding flux is taken as the contrast limit at that separation. This completeness-based contrast therefore provides an empirical estimate of the detection sensitivity of the algorithm, without relying on assumptions about the underlying noise distribution.

    \section{Results}
    \label{sec:results}

NA-SODINN was run separately on each preprocessed ADI sequence of the F150 sample. This involved, for every sequence, first identifying noise regimes (\cref{sec:noise_regimes}), and then, for each regime, generating training data (\cref{sec:training_set}), training a Conv-LSTM binary classifier (\cref{sec:model_training}), and performing inference (\cref{sec:inference}), ultimately producing both confidence and \fOneScore maps (\cref{sec:f1_threshold}). Hereafter, we refer to both of them as detection maps. For targets with multiple observational epochs, each epoch was processed independently, meaning the full NA-SODINN pipeline was applied separately to each one. Similarly, H2 and H3 filter sequences for the same target were treated as distinct datasets.

A valid detection was defined as an ensemble of one or more connected pixels \citep{Cantalloube_2020} appearing in the confidence map, passing the threshold mask, and therefore also present in the \fOneScore binary map. Applying this criterion within our reduced field of view, NA-SODINN identified a total of 30 point-like candidate sources across 19 targets in the full survey, excluding confirmed gravitationally bound companions and disks. Some of these candidates had already been reported in previous studies, while others are newly identified here. To facilitate their analysis, this section is organized as follows. In \cref{sec:previous_discoveries}, we review previously reported sources recovered by NA-SODINN, including confirmed substellar companions, detections previously classified as background contaminants or ambiguous cases in the literature, and known disks. In \cref{sec:new_candidates}, we present the candidate sources not reported by earlier studies. Finally, \cref{sec:sensitivity_limits} concludes with an evaluation of the sensitivity limits reached by NA-SODINN for the F150 sample. 

Two appendices complement this section. 
A selection of NA-SODINN detection maps is provided in \cref{app: detection maps}, including some of the confirmed recovered planets and disks, as well as the candidate sources newly identified in this work.
\Cref{app:characterization_candidates} summarizes the main properties of the target stars around which NA-SODINN identified new candidates (\cref{Table:targets_detections}), together with the characterization of the candidates themselves (\cref{Table:targets_detections_characterization}).

\subsection{Previously reported discoveries\label{sec:previous_discoveries}}

\subsubsection{Known substellar companions\label{sec:known_substellar_companions}}
All confirmed companions present in the F150 sample were detected by NA-SODINN with high confidence scores. This includes the seven brown dwarfs PZ Tel B \citep{Biller_2010}, HD 115470 B \citep{Cheetham_2018}, $\eta$ Tel B \citep{Lowrance_2000}, CD-35 2722 B \citep{Wahhaj_2011}, HD 143567 B \citep{Lafreniere_2011}, HD 206893 B \citep{Milli_2017_discovery}, and CD-52 381 B \citep{Chauvin_2005_bw}; as well as the nine planetary-mass companions 51 Eri b \citep{Macintosh_2015}, $\beta$ Pic b \citep{Lagrange2009AProbable}, HD 95086 b \citep{Rameau_2013_HD95086}, HR~8799 c, d, and e \citep{Marois_2008_HR8799, Marois_2010_HR8799}, HD 116434 b \citep{Chauvin_2017_HIP65426}, GJ 504 b \citep{Kuzuhara_2022}, and AB Pic b \citep{Chauvin_2005}.

For illustrative purposes, \cref{fig:detection_maps_known_exoplanets} presents four examples of these detections recovered in both the H2 and H3 filters. For instance, \cref{fig:detmaps_knownComp_HD218396} shows the case of HR~8799. We observe that the three giant planets HR~8799~c, d, and e are recovered by NA-SODINN, while planet~b is not detected as it falls outside our reduced $1\farcs2$ field of view. The presence of low-mass inner companion interior to HR~8799~e could help stabilize the system dynamically \citep{Gozdziewski_2014}, although deep SPHERE observations have ruled out planets more massive than $\sim3$-4~$M_{\mathrm{Jup}}$ in the 7.5-10~au range \citep{Wahhaj_2021_HR8799_fifthplanet}. In \cref{fig:detmaps_knownComp_HD218396}, aside from the detection of planets c, d, and e, the NA-SODINN confidence maps also reveal a few other blobs that did not pass the threshold mask. Notably, in the H3 filter, a cluster of pixels with $\sim0.8$ confidence appears at a slightly smaller angular separation than planet~e. Although this signal could be considered significant, we opted not to include it in our posterior analysis due to its non-detection in the H2 filter and the absence of a corresponding signal in other epochs. For the systems HIP 65426, PZ Tel, and $\beta$ Pic, whose NA-SODINN detection maps are shown in \cref{fig:detmaps_knownComp_HIP65425,fig:detmaps_knownComp_HIP92680,fig:detmaps_knownComp_HIP27321}, the known substellar companions are robustly recovered within the considered field of view. No additional significant point-like sources are identified in the corresponding confidence or binary maps. The same result is obtained for the other analogous systems in the F150 sample, including the seven brown dwarf hosts and the three additional exoplanet host stars.

\subsubsection{Background contaminants and ambiguous cases}
\label{sec:bck_reported_candidates}

Seventeen NA-SODINN detections correspond to candidate point sources previously reported by \citet{Langlois_2021_shine}, \citet{Chomez_2025_shine}, and \citet{Sabalbal_2025}. In those studies, these sources were ultimately classified either as background contaminants or as ambiguous cases. Five of them were identified as background contaminants, around HD175726, HD151726, HD148055, and HD107821, based on their position in color-magnitude diagrams and/or rejection via the common proper-motion test. Six others were flagged as ambiguous, around HD113457, HD107301, OUPup, and HD326277, typically because the astrometric and/or photometric measurements were not sufficiently reliable for a robust assessment. The remaining six are found in systems with additional noteworthy structures: two lie beyond the debris disk of HD115600 (\cref{detmaps_knownDisks_HD115600}), and four are found in systems for which we also report new candidates (\cref{sec:new_candidates}): three around HD168210 (\cref{detmaps_target_b_HD168210}) and one around HD~164249 (\cref{detmaps_target_f_HD164249A}).

\subsubsection{Known debris disks\label{sec:known_disks}}
The F150 sample includes twelve confirmed debris disks. Nine of these were previously known, corresponding to the targets: $\beta$ Pic \citep{Lagrange_2019}, NZ Lup \citep{Boccaletti_2019}, HD 115600 \citep{Gibbs_2019}, HD 197481 \citep{Boccaletti_2018}, HD 15115 \citep{Engler_2019}, HD 61005 \citep{Olofsson_2016}, HD 377 \citep{Choquet_2016}, CE Ant \citep{Olofsson_2018}, and HD 109573 \citep{Milli_2017}. The remaining three, HD 160305 \citep{Perrot_2019}, HD 106906 \citep{Lagrange_2016_discovery_disk}, and HD 131835 \citep{Feldt_2017_discovery}, were discovered within the SHINE survey. These disks were typically found around young stars, with a median stellar age of 24~Myr, and are most often observed at high inclinations, with a median value of $i = 82$\textdegree{} \citep{Langlois_2021_shine}, except for the case of CE Ant, which was detected at low inclination through polarimetric differential imaging \citep{Olofsson_2018}. This prevalence of high-inclination detections is likely due to a selection bias introduced by ADI, which favors the detection of edge-on, optically thin disks through enhanced forward scattering and improved contrast performance.

Given our smaller field of view and the fact that the NA-SODINN neural network is trained to identify point-like sources rather than extended sources, we recovered, as shown in \cref{fig:known_disks}, partial disk morphologies in only three of these systems. For example, \cref{detmaps_knownDisks_HD109573} shows the case of HD~109573, an A0V star with an estimated age of $10 \pm 3$ Myr \citep{Bell_2015} and at a distance of $71.91 \pm 0.70$ pc \citep{Gaia_collabortion_2022}, that harbors one of the disks with the highest fractional luminosity, shaped as a thin ring of semi-major axis $\sim 77$ au inclined by $\sim 76$\textdegree. We can observe that the NA-SODINN recovery of the disk morphology in both filters is surprisingly good, with very high confidence scores and no additional spurious detections. We attribute this unexpected result to the interplay between the S/N curves included during training (see \cref{sec:training_set}) and their correlation with the patch sequence. Although the $c_{+}$ patch sequences contained only centered point-like source injections of varying fluxes and different noise levels, and no extended structures, the exceptionally high flux of HD109573’s disk produced strong S/N values across the corresponding S/N curves. This allowed the network to almost recover the full disk despite its lack of explicit training on extended features.

\Cref{detmaps_knownDisks_HD115600} presents the case of HD~115600, an F2V/F3V star with an estimated age of $\sim15$~Myr and located at $109.6 \pm 0.5$~pc \citep{Gaia_collabortion_2022}. This system harbors a smaller debris disk with a semi-major axis of $\sim48$~au and an inclination of $\sim79$\textdegree. In contrast to HD109573, we observe that NA-SODINN struggles to recover the complete morphology of this disk, leaving significant gaps along its structure. Here, part of the disk lies within the speckle-dominated regime. As a result, despite the disk being intrinsically bright, the lower and more irregular S/N response in this area prevents the network from confidently identifying the entire ring, limiting the recovery to fragmented portions of the disk.

Finally, \cref{detmaps_knownDisks_HD197481} shows HD~197481 (AU Mic), an M3IVe star at $9.725 \pm 0.005$~pc that hosts two confirmed hot Neptunes \citep{Plavchan_2020, Martioli_2020} and possibly a third planet \citep{Wittrock_2023}. Its edge-on disk, one of the largest and brightest known, extends from 40 to 120~au. In this case, NA-SODINN detects only a minor portion of the disk and no substellar candidates.

\subsection{New candidates}
\label{sec:new_candidates}

In addition to the known companions, our analysis revealed 13 new candidates around 11 stars not previously reported in the literature. \Cref{Table:targets_detections} provides the main properties of these stars. We group these detections into two categories: 10 candidates detected in both H2 and H3 bands, including Smethells 20 (three candidates), HD 168210, $\pi$ Ara, CD-51 10924, $\beta$ Leo, HD 164249, CD-57 1054, and CE Ant; and 3 candidates detected in only one of the two bands, namely CD-48, HD 55279, and HD 219482. \Cref{fig:detmaps_H23_newCandidates,fig:detmaps_Honly_newCandidates} show the NA-SODINN detection maps for both groups, respectively.

The relative astrometry and photometry of these new candidates were estimated using the negative fake companion method \citep[NEGFC,][]{Wertz_2017}, as implemented in the \texttt{VIP} package \citep{Gomez_2017, Christiaens2023VIP}. Importantly, this characterization step is independent of NA-SODINN: while NA-SODINN was used to identify the candidate locations, the estimation of their astrometry and photometry relies solely on the signal present in the final ADI-PCA post-processed frame. Therefore, this procedure is only applicable to candidates that remain sufficiently visible after ADI-PCA processing. In practice, the method consists of injecting a fake companion with negative flux at the approximate location of the detected candidate in the ADI sequence. If the injected negative source has the correct position and flux, it will cancel the candidate signal in the ADI-PCA post-processed frame. To achieve this, the separation, position angle, and flux of the negative companion were adjusted through an iterative procedure that involved two steps. First, a Nelder-Mead simplex algorithm was used to obtain initial estimates. Second, these estimates were refined, and statistical uncertainties were obtained by sampling the posterior distributions with an MCMC sampler. \Cref{Table:targets_detections_characterization} presents this characterization for all the new candidates.

\begin{figure}[]
    \centering
    \includegraphics[width=0.99\linewidth]{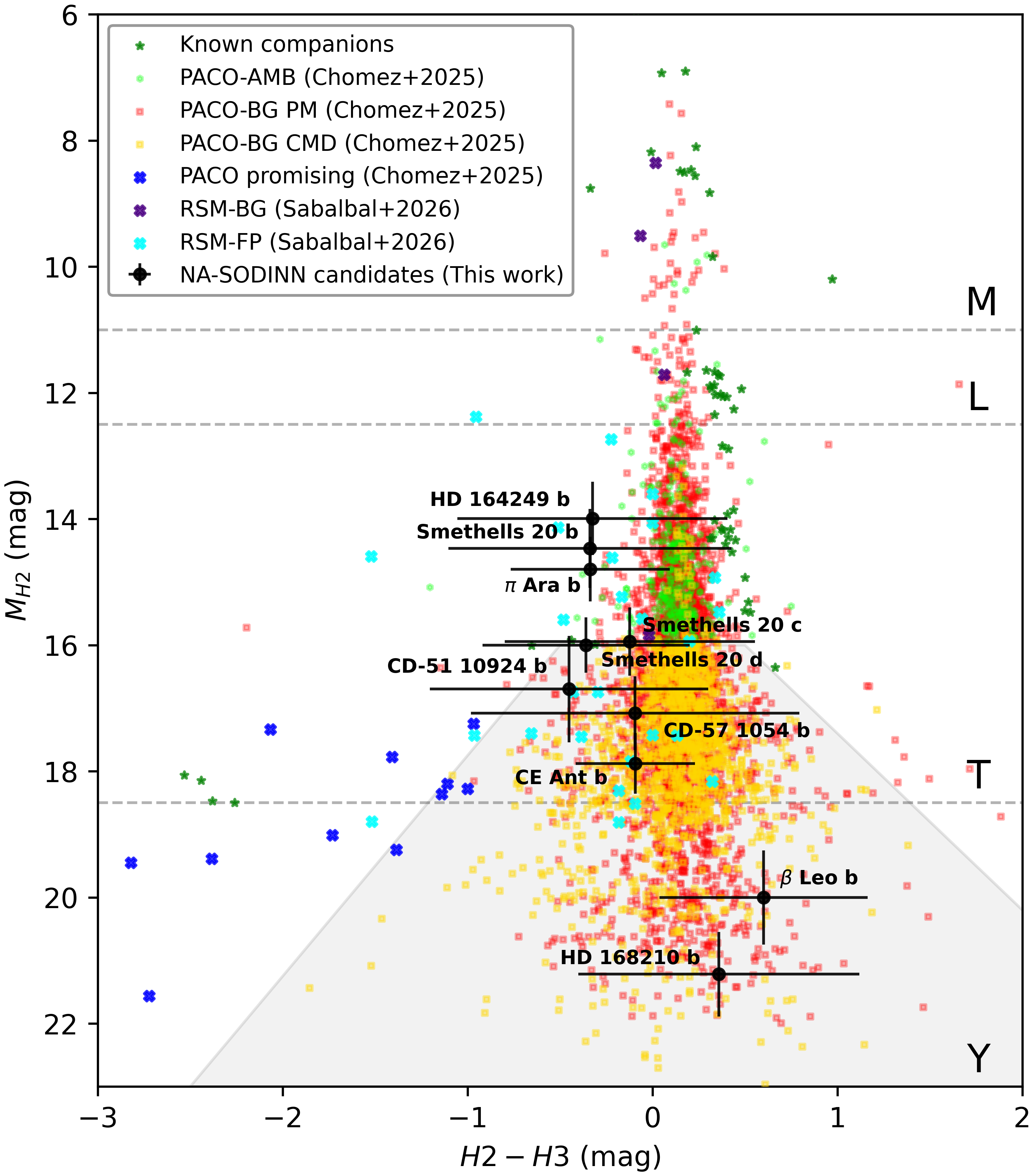}
    \caption{H2--H3 color--magnitude diagram of the 10 NA-SODINN candidates detected in both bands (black circles with photometric uncertainties and labels). Previously known SHINE companions are shown as green stars. Candidate sources reported by \citet{Chomez_2025_shine} and \citet{Sabalbal_2025} are overplotted and colour-coded according to their final classification: ambiguous (PACO-AMB; light green), background contaminants rejected through proper-motion analysis (PACO-BG PM; red), background contaminants rejected through the CMD (PACO-BG CMD; yellow), promising candidates from PACO (PACO promising; blue), additional background sources identified by RSM (RSM-BG; purple), and RSM false positives (RSM-FP; aqua). The grey shaded region indicates the empirical background-exclusion zone defined by \citet{Chomez_2025_shine}, while horizontal dashed lines and labels mark approximate spectral-type transitions.}
    \label{fig:cmd}
\end{figure}

After deriving the astrometry and photometry, we placed the 10 candidates detected in both bands on a color–magnitude diagram (CMD; \cref{fig:cmd}; see also \citealt{Bonnefoy_2018}) to evaluate their nature. The CMD is a well-established diagnostic for distinguishing substellar companions from background contaminants, particularly beyond the L/T transition where silicate and iron clouds dissipate and methane absorption dominates, producing the characteristic blueward shift of T dwarfs. Following \citet{Chomez_2025_shine}, we also show the empirically defined exclusion region that traces the locus of the dominant SHINE contaminants (field K- and M-type stars). For context, \cref{fig:cmd} additionally includes the population of previously confirmed SHINE companions, as well as the candidate detections reported by \citet{Chomez_2025_shine} and \citet{Sabalbal_2025}, using PACO and RSM algorithms, respectively, color-coded by their final classifications. Note that the RSM signals presented in this diagram represent an updated version of the results published in \citet{Sabalbal_2025}, with signal characterization now performed using NEGFC.

Based on their photometric and astrometric properties, we preliminarily classify our candidates into three categories: (i) \textit{background}, when the source is unlikely to be physically bound to the host star and is most probably a field star projected by chance within the field of view (see \cref{subsec:background}); (ii) \textit{ambiguous}, when the available measurements do not allow a clear distinction between a background contaminant and a bound companion, often due to large uncertainties or conflicting indicators (see \cref{subsec:ambiguous}); and (iii) \textit{promising}, when the source properties are consistent with expectations for substellar companions, making it a plausible bound object and, therefore, a good target for follow-up observations (see \cref{subsec:promising}). The three candidates detected in only one band cannot be placed on the H2–H3 CMD and are therefore left \textit{undefined} at this stage. This classification is also highlighted in  \cref{Table:targets_detections_characterization} and in NA-SODINN detection maps of \cref{fig:detmaps_H23_newCandidates,fig:detmaps_Honly_newCandidates}.

\subsubsection{Background}
\label{subsec:background}
Substellar candidates labeled as background contaminants are those that fall within the exclusion region in the CMD of \cref{fig:cmd}. We identify five clear cases: HD168210 b, $\beta$ Leo b, CE Ant b, CD-57 1054 b, and CD-51 10924 b. 

The HD 168210 system \citep{Perugini_2021} has been the subject of several HCI surveys, including the VLT/NaCo Large Program \citep{Chauvin_2015} and the International Deep Planet Survey \citep{Galicher_2016}, which reported multiple candidate sources that were later confirmed as background stars. In our analysis, NA-SODINN recovers four sources (see \cref{detmaps_target_b_HD168210}): three of them match previously reported background contaminants, while an additional candidate, HD168210b, is detected with very high confidence ($\sim0.98$) at a projected separation of $\sim569\pm20$ mas. In the CMD, this source falls well inside both empirical exclusion regions and overlaps the locus of background stars.

A similar situation is found in $\beta$ Leo (HIP 57632), a bright A3V star with a prominent infrared excess from circumstellar dust \citep[e.g.,][]{Cote_1987}. NA-SODINN recovers a single candidate at $\sim373\pm24$ mas (\cref{detmaps_target_e_HIP57632}), detected with high confidence ($\sim0.97$). However, its photometry lies squarely within the exclusion region in the CMD, favoring a background interpretation. Moreover, $\beta$~Leo has a very large proper motion ($\sim 500$ mas/yr), so a stationary background source would be expected to show a substantial relative displacement with respect to the star between epochs and could therefore appear at noticeably different (and potentially larger) separations in earlier high-contrast imaging datasets. However, proper-motion vetting and dedicated follow-up observations are beyond the scope of this paper and are left for future work.

CE~Ant (TWA 7) is an M2 star and a member of the TW Hydra young association \citep{Webb_1999}. It hosts a nearly pole-on debris disk with multiple rings \citep{Choquet_2016, Ren_2021}, and recently \citet{Lagrange_2025} reported the direct imaging of a cold, sub-Jovian planet at $\sim1$\farcs5 with JWST/MIRI—well outside our reduced field of view. Within our data, NA-SODINN recovers a point-like source at $\sim328\pm28$ mas (\cref{detmaps_target_g_CEAnt}), whose photometry lies securely within the exclusion regions of the CMD, again pointing to a background contaminant.

Among the young moving group members, CD-57 1054 (HIP 23309) stands out as a nearby M0 dwarf associated with the $\beta$~Pic moving group \citep{Lee_2024}. This star has been the focus of several exoplanet searches, including SHINE; however, no companions have yet been confirmed. Recently, \citet{Gollotti_2024} identified two dips in TESS light curves that suggest a possible planet, but follow-up HARPS radial velocities remain inconclusive. In this context, the NA-SODINN detection maps (\cref{detmaps_target_g_HIP23309}) reveal several faint features, of which only one meets the threshold mask criterion. The candidate, at $\sim555\pm30$ mas, lies inside the inner exclusion region of the CMD, close to its boundary, with colors similar to CE~Ant~b. Its CMD placement favors a background interpretation.

The field M-dwarf CD-51 10924 (GJ 676 A) is part of a wide binary and hosts one of the most diverse planetary systems known around such stars. \citet{Forveille_2011} first identified a massive companion (planet b) through radial-velocity monitoring and detected an additional long-term drift that could not be explained by the stellar companion GJ 676 B, suggesting the presence of a second giant planet c; this was later confirmed by \citet{Anglada_2012}, who also discovered two inner super-Earths (planets d and e) around GJ 676 A. In our NA-SODINN detection maps (\cref{detmaps_target_d_HIP85647}), a point-like source is clearly recovered at a separation of $\sim287\pm27$ mas in the single F150 epoch available for this target. In the CMD (\cref{fig:cmd}), the source lies within the exclusion region. Astrometrically, the candidate would fall between planets b and c if bound, yet its brightness is inconsistent with such a scenario, favoring a background interpretation.

\subsubsection{Ambiguous}
\label{subsec:ambiguous}

Substellar candidates labeled as ambiguous are those that, within their photometric uncertainties, fall along the boundaries of the exclusion regions in \cref{fig:cmd}. We identify two such cases: Smethells 20 c and d.

The young star Smethells 20 (or TYC 9073-762-1) is a member of the $\beta$ Pictoris moving group \citep{Binks_2017, Shan_2017}, and to date, no substellar companions have been confirmed in this system. Previous HCI surveys, including SHINE, as well as the MASSIVE program with VLT/NaCo \citep{Lannier_2017} and the Gemini/NICI Planet-Finding Campaign \citep{Biller_2013}, reported additional candidate sources at wider separations, outside our reduced field of view, which were classified as background objects. However, as shown in \cref{detmaps_target_a_2MASSJ1846}, NA-SODINN identifies multiple candidate companions in the only epoch available in the F150 sample for this star. The two outer candidates, Smethells 20 c and d, at angular separations of $\sim450\pm30$ mas and $\sim680\pm28$ mas (see \cref{Table:targets_detections_characterization}), respectively, lie at the edge of the exclusion region in the CMD (\cref{fig:cmd}). Their near-neutral colors are consistent with the L/T transition, with dust-free to moderately dusty atmospheres. If physically bound, their photometry is consistent with $\sim6$–8 $M_{\mathrm{Jup}}$ companions according to the AMES-DUSTY evolutionary model, although their uncertainties also place them within the background locus, leaving their nature ambiguous.

\subsubsection{Promising}
\label{subsec:promising}

Substellar candidates classified as promising are those that lie entirely outside the exclusion regions. We identify three such cases: Smethells 20 b, $\pi$ Ara b, and HD 164249 b.

Unlike Smethells 20 c and d, the inner candidate (\cref{detmaps_target_a_2MASSJ1846}) detected at $\sim220$ mas from the star (\cref{Table:targets_detections_characterization}), designated Smethells 20 b, lies outside the exclusion regions in \cref{fig:cmd}. Within the photometric uncertainties, its position appears more consistent with $\sim9$–11 $M_{\mathrm{Jup}}$ companions according to the AMES-DUSTY evolutionary model, while its relatively H2 - H3 color suggests the onset of CH$_4$ absorption in the near-IR, characteristic of young early-T substellar objects. No additional SPHERE epochs are currently available for an astrometric check of common proper motion. We also identified several archival NACO observations of this target, including a relatively deep $L'$ pupil-tracking sequence, although the small angular separation of the candidate is likely to limit the constraining power of these data. 

\Cref{detmaps_target_c_HIP86305} presents the detection maps of the star $\pi$~Ara \citep{Zakhozhay_2022}. In this case, NA-SODINN identifies a faint source at  $\sim 900 \pm 25$ mas. To date, no bound companions have been confirmed around this system. Previous HCI, including HST/ACS coronagraphy \citep{Doering_2007}, the SHINE survey, and the Gemini Planet-Finding Campaign \citep{Wahhaj_2013}, likewise reported only background sources or non-detections. A second epoch, obtained on 2019-05-19 and available through the HC-DC, was also analyzed with NA-SODINN, but the candidate detected in the F150 epoch was not recovered.  

The F6 star HD 164249 (HIP 88399) is a member of the $\beta$ Pictoris moving group and hosts a known M2 stellar companion, HD 164249 B. An infrared excess, first detected with WISE \citep{Wright_2010}, Spitzer \citep{Chen_2014}, and Herschel \citep{Eiroa_2013}, revealed the presence of a debris disk that was later spatially resolved with ALMA \citep{Pawellek_2021}. Despite the absence of a direct detection in multiple SPHERE epochs \citep{Mesa_2022}, the star exhibits a significant proper motion anomaly indicative of an additional low-mass companion. Recent analyzes suggest the presence of a Jupiter-like planet with a mass of $\sim4$–5 $M_{\mathrm{Jup}}$ at a semi-major axis of $\sim7$ au \citep{Gratton_2024, Gonzalez_Payo_2024, Lagrange_2025_searching}, although this companion remains unconfirmed. The NA-SODINN detection maps reveal two point sources (\cref{detmaps_target_f_HD164249A}): one located close to the outer edge of the field of view, which was also reported and classified as a background contaminant by \citet{Langlois_2021_shine}, and another faint source at $\sim330$ mas ($\sim16$ au). Although this latter detection does not coincide with the separation predicted for the putative low-mass companion inferred from astrometry, its position in the CMD (\cref{fig:cmd}) overlaps with the locus of planetary-mass objects, making it an interesting candidate. However, five additional SPHERE epochs available through the HC-DC and obtained between 2015 and 2019 were also analyzed with NA-SODINN, and the same candidate was not recovered in any of them.

\begin{figure}[!t]
    \centering
    \includegraphics[width=0.99\linewidth]{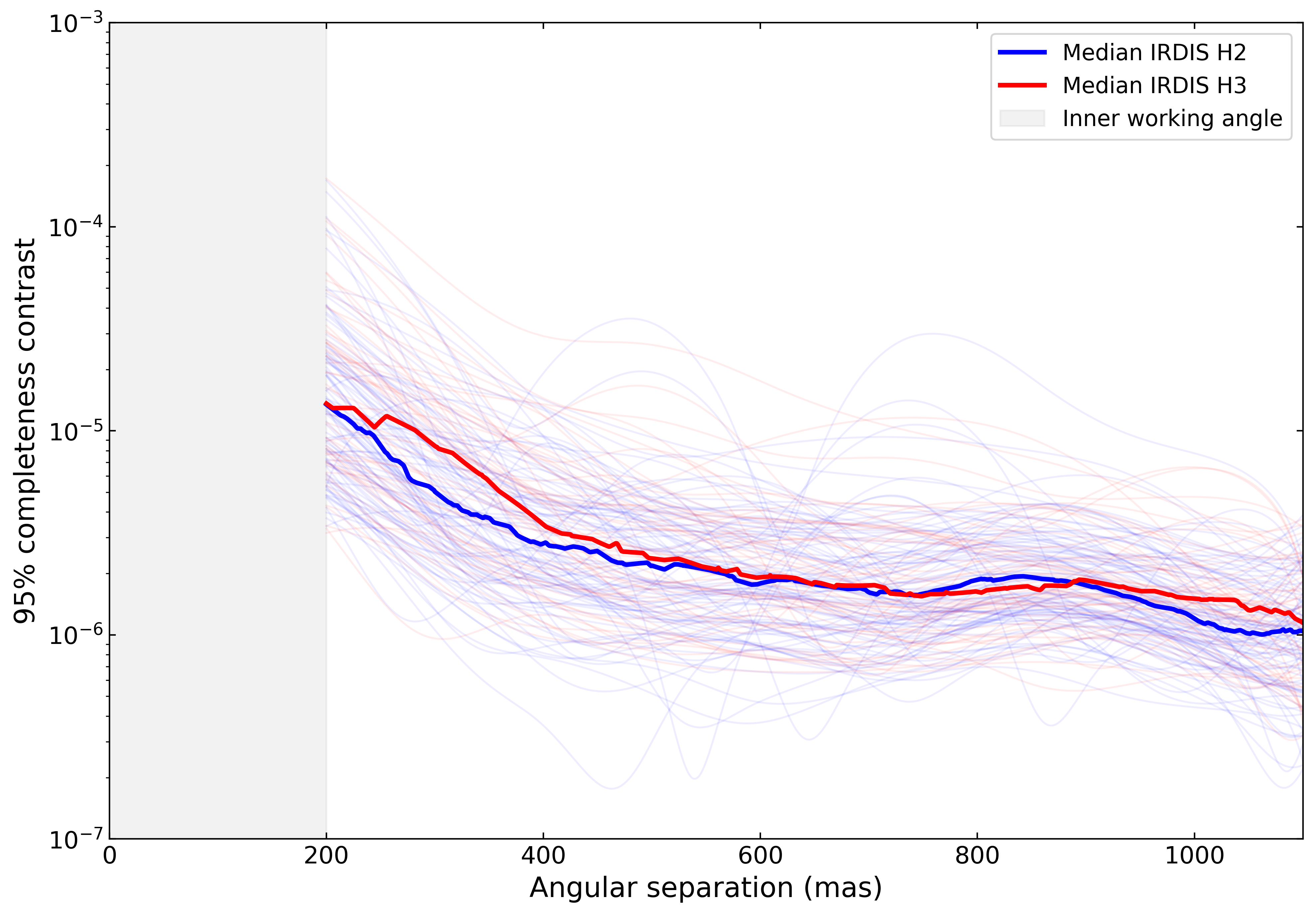}
    \caption{NA-SODINN contrast curves, based on 95\% completeness, computed for the set of ADI sequences in the F150 sample in both H2 (blue) and H3 (red) pass band filters. Thin curves represent individual sequences, while the thick curves correspond to the median contrast curve in each band. The gray shaded region marks the inner working angle.}
    \label{fig:contrast_curves}
\end{figure}

\subsection{Sensitivity limits\label{sec:sensitivity_limits}}

As a final step in the reprocessing of the F150 sample with NA-SODINN, we computed 95\% completeness contrast curves for each target following the procedure described in \cref{sec:contrast_curves}. The resulting curves in the IRDIS H2 and H3 filters are shown in \cref{fig:contrast_curves}. Beyond the inner working angle ($\sim200$~mas), the median 95\% completeness contrast improves rapidly, reaching $\sim10^{-5}$ at $\sim250$~mas in both filters, with slightly deeper limits in H2. At $\sim1$\arcsec, the median contrast reaches values close to $\sim10^{-6}$ in both H2 and H3.

As discussed in \cref{sec:contrast_curves}, these limits are not directly equivalent to standard $5\sigma$ contrast curves. Nevertheless, a comparison with previous IRDIS results obtained with PCA \citep{Langlois_2021_shine}, PACO \citep{Chomez_2025_shine}, and RSM \citep{Sabalbal_2025} remains informative at the survey level. Relative to those methods, the NA-SODINN contrast curves exhibit a markedly reduced target-to-target dispersion over the full separation range. In terms of sensitivity gain, NA-SODINN appears broadly comparable to RSM within the reduced field of view explored here, and therefore also to PACO, given the similar performance reported for RSM and PACO in that separation regime.

    \section{Conclusions}
    \label{sec:conclusions}

In this work, we presented a reanalysis of the F150 sample \citep{Desidera_2021_shine, Langlois_2021_shine, Vigan_2021_shine} from the SPHERE High-contrast Imaging survey for Exoplanets \citep[SHINE, ][]{Chauvin_2017_shine} using the recently developed deep-learning detection algorithm NA-SODINN \citep{Cantero_2023}. Given NA-SODINN’s proven ability to discriminate faint substellar companions from residual speckle noise in ADI–PCA processed frames, our objective was to search for additional companion candidates in this first public SHINE release that may have been missed by previous analyzes.

To achieve this, we introduced two new modifications to the algorithm. First, since NA-SODINN produces confidence maps with uncalibrated probabilities rather than standard S/N maps, we implemented a principled detection criterion based on an \fOneScore-score-driven thresholding strategy. This procedure masks the confidence map annulus by annulus, providing an objective criterion for companion detection. Second, to compute detection limits, we adopted contrast curves based on 95\% completeness instead of the classical $5\sigma$ criterion, and we optimized NA-SODINN pipeline to calculate these curves for each target in the survey.

Based on these improvements and within the broader context of recent SHINE re-analyzes, we reprocessed the full F150 sample on a sequence-by-sequence basis, establishing NA-SODINN as a competitive deep-learning alternative to existing approaches, such as the original PCA-based processing by \citet{Langlois_2021_shine}, the PACO-based full-survey reprocessing by \citet{Chomez_2025_shine}, and the RSM-based reprocessing of the F150 sample by \citet{Sabalbal_2025}. NA-SODINN recovered all previously known companions in the survey, detected part of the circumstellar disks, and identified 30 additional substellar candidates. Of these, 17 had already been reported in the aforementioned studies and classified as either background contaminants or ambiguous cases. The remaining 13 sources, found around 11 stars, are reported here for the first time. Among them, three were detected in only one of the two IRDIS H23 spectral bands, while 10 were identified in both. For these 10 sources, we used the H2-H3 color-magnitude diagram to obtain a first-order assessment of their nature, classifying them into three groups: five background contaminants, two ambiguous cases, and three photometrically promising companion candidates around the stars Smethells, $\pi$ Ara, and HD 164249. However, only the candidate Smethells 20 b remains as promising based on the currently available data, while the non-recovery of $\pi$ Ara b and HD 164249 b in additional SPHERE epochs weakens their candidacy.

This study represents the first application of NA-SODINN to real data from a large high-contrast imaging survey. It therefore provides a benchmark for validating the method on survey observations and highlights clear directions for improvement. First, the per-sequence reduction strategy proved effective, but at the expense of substantial computational cost: processing a single ADI sequence in one band with NA-SODINN typically requires 10-30 hours in its current CPU multiprocessing implementation. This cost depends mainly on the number of frames in the sequence and, in particular, on the computation of the contrast curve, which is the dominant bottleneck. Reducing this cost will likely require the adaptation of alternative computational strategies. For example, clustering-based approaches \citep[e.g.,][]{Dahlqvist_2002_shards, Sabalbal_2025} could be used to group ADI sequences with similar characteristics (e.g., atmospheric conditions) and enable the training of a single NA-SODINN model per cluster rather than training a separate model for each sequence.
Second, the presented reprocessing approach can be extended beyond the F150 sample. A natural next step would be to reanalyze the full SHINE survey, enabling a more comprehensive statistical assessment of the companion population, and to adapt the pipeline to other ground-based HCI programs such as Gemini-GPIES \citep{Nielsen2019TheGemini}, NACO-ISPY \citep{Launhardt_2020}, and SPHERE-BEAST \citep{Janson_2021_}.

The field of HCI is rapidly evolving with the development of powerful post-processing techniques based on diverse principles for speckle modeling, speckle suppression, and point-like source detection. Applying such algorithms to archival surveys represents a cost-effective strategy to maximize their scientific return, as demonstrated in this work. Our results illustrate the potential of deep-learning approaches not only to recover known companions but also to reveal previously undetected candidates, underscoring the importance of re-examining existing datasets with modern tools while preparing for the next generation of HCI instruments such as SAXO+ \citep{Galland_2024} and GPI 2.0 \citep{Chilcote_2024}.

\begin{acknowledgements}
      This work has been carried out within the framework of the National Center of Competence in Research PlanetS supported by the Swiss National Science Foundation. The authors would like to thank the open-source Python scientific community, and in particular the developers of the Keras deep learning library \citep{Abadi_2015} and the VIP high-contrast imaging package \citep{Gomez_2017, Christiaens2023VIP}, which were key to this work. We also extend our gratitude to the Exoplanet Imaging Data Challenge community \citep{Cantalloube_2020} for fostering a collaborative environment and contributing valuable tools and benchmarks to the field of HCI. We thank the entire SPHERE consortium for their long-standing efforts in the design, operation, and continued support of the SPHERE instrument, which made this survey possible. This work has made use of the High Contrast Data Centre, jointly operated by OSUG/IPAG (Grenoble),PYTHEAS/LAM/CeSAM (Marseille), OCA/Lagrange (Nice), Observatoire de Paris/LESIA (Paris), and Observatoire de Lyon/CRAL, and is supported by a grant from Labex OSUG@2020(Investissements d'avenir -- ANR10 LABX56). This project has received funding from the European Research Council (ERC) under the European Union’s Horizon 2020 research and innovation program (grant agreement No 819155), and from the Wallonia-Brussels Federation (grant for Concerted Research Actions).
\end{acknowledgements}

\begin{appendix}

        \clearpage
        \section{The \fOneScore-score\label{app:F1_score}}
        
        In a detection problem, such as exoplanet detection with NA-SODINN, each candidate can be assigned to one of four categories: true positive (TP), false positive (FP), or false negative (FN). These quantities form the basis for standard performance metrics such as \emph{precision}, \emph{recall} (also named \emph{true positive rate}, TPR), and the \emph{\fOneScore-score}.
        
        In this work, a true positive corresponds to a detected injected planetary signal, while a false positive corresponds to a spurious detection caused by residual speckle noise or image artifacts. We assume that the analyzed data sets do not contain real astrophysical companions, such as genuine exoplanets or background stars, so that all positive detections can be unambiguously classified as either true or false.
        
        \emph{Precision} measures the fraction of positive detections that are correct, and is defined as
        \begin{equation}
            \label{eq:precision}
            \mathrm{Precision} = \frac{\mathrm{TP}}{\mathrm{TP} + \mathrm{FP}}.
        \end{equation}
        It ranges from 0 to 1. A value close to 1 indicates that most detections correspond to injected planetary signals, with only a small number of false positives. However, optimizing only for high precision may lead to an overly conservative strategy in which faint or marginal signals are missed.
        
        \emph{Recall}, on the other hand, measures the fraction of injected signals that are successfully recovered:
        \begin{equation}
            \label{eq:recall}
            \mathrm{Recall} = \frac{\mathrm{TP}}{\mathrm{TP} + \mathrm{FN}}.
        \end{equation}
        A high recall indicates that most injected companions present in the data are detected. In practice, however, increasing recall often comes at the expense of lower precision, since a more permissive detection criterion generally yields more false positives.
        
        Because precision and recall are intrinsically coupled, the \emph{\fOneScore-score}, defined as their harmonic mean, provides a convenient summary of the trade-off between the values of both quantities\footnote{Note that, while \fOneScore is a good trade-off between the values of the precision and the recall, \citet{Pierard2026What} have shown that \fOneScore is not necessarily a good trade-off between the ranks obtained for the precision and the recall. In other words, \fOneScore is adequate for optimizing a method, as we do, but not necessarily for comparing algorithms.}:
        \begin{equation}
            \label{eq:F1-score}
            F_1 =
            \frac{2 \cdot \mathrm{Precision} \cdot \mathrm{Recall}}
            {\mathrm{Precision} + \mathrm{Recall}}
            =
            \frac{2 \cdot \mathrm{TP}}
            {2 \cdot \mathrm{TP} + \mathrm{FP} + \mathrm{FN}}.
        \end{equation}
        
        A high \fOneScore-score indicates that the method achieves both high completeness and high reliability, that is, it recovers a large fraction of injected signals while keeping the number of false detections low. In the context of NA-SODINN, optimizing the detection threshold using the \fOneScore-score provides a practical compromise between precision and recall. This is particularly relevant in HCI, where the noise properties vary with angular separation and a single global threshold may not be optimal. By adapting the threshold on an annulus-by-annulus basis, the method yields a more robust and interpretable detection strategy.

    \section{NA-SODINN detection maps\label{app: detection maps}}
\begin{figure*}
        \centering
        \begin{subfigure}[b]{\textwidth}
            \includegraphics[width=0.99\textwidth]{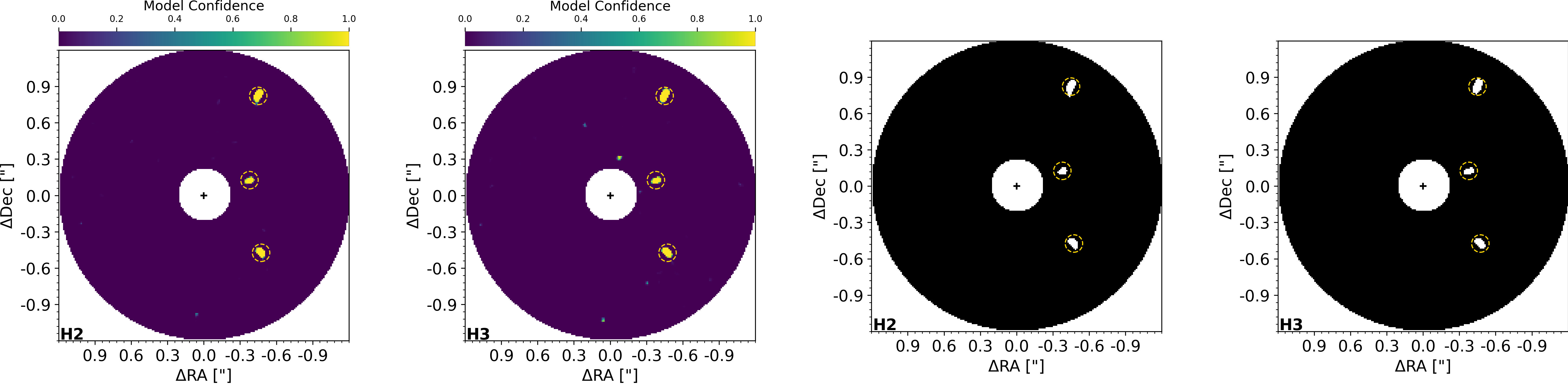} 
            \caption{Target: \textbf{HD 218396}\hspace{0.3cm}Observing date: \textbf{2017-06-14}\hspace{0.3cm}\label{fig:detmaps_knownComp_HD218396}}
        \end{subfigure}\vspace{0.5cm}
        \begin{subfigure}[b]{\textwidth}
            \includegraphics[width=0.99\textwidth]{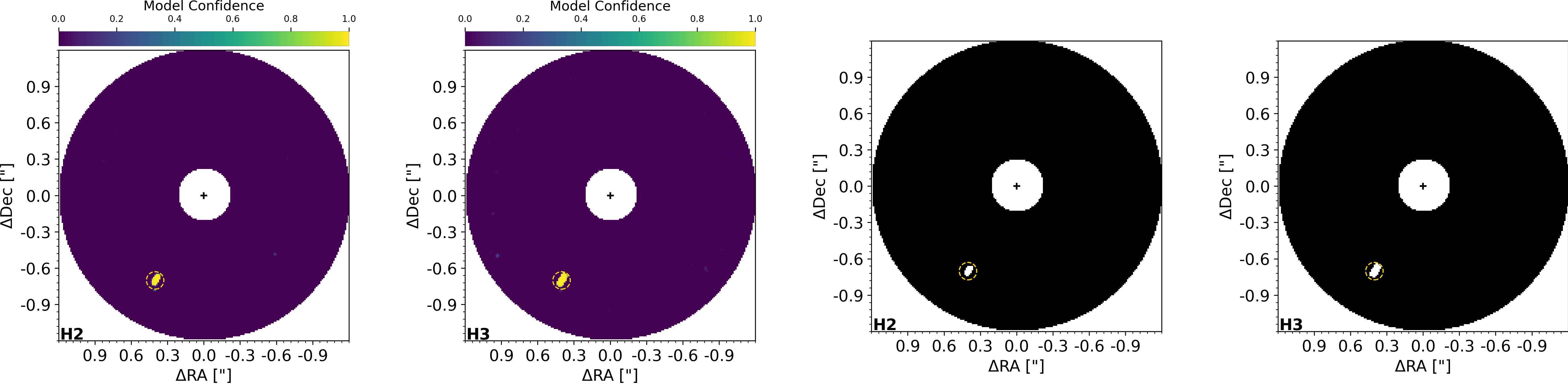} 
              \caption{Target: \textbf{HD 116434}\hspace{0.3cm}Observing date: \textbf{2016-06-26}\hspace{0.3cm}\label{fig:detmaps_knownComp_HIP65425}}
        \end{subfigure}\vspace{0.5cm}
        \begin{subfigure}[b]{\textwidth}
            \includegraphics[width=0.99\textwidth]{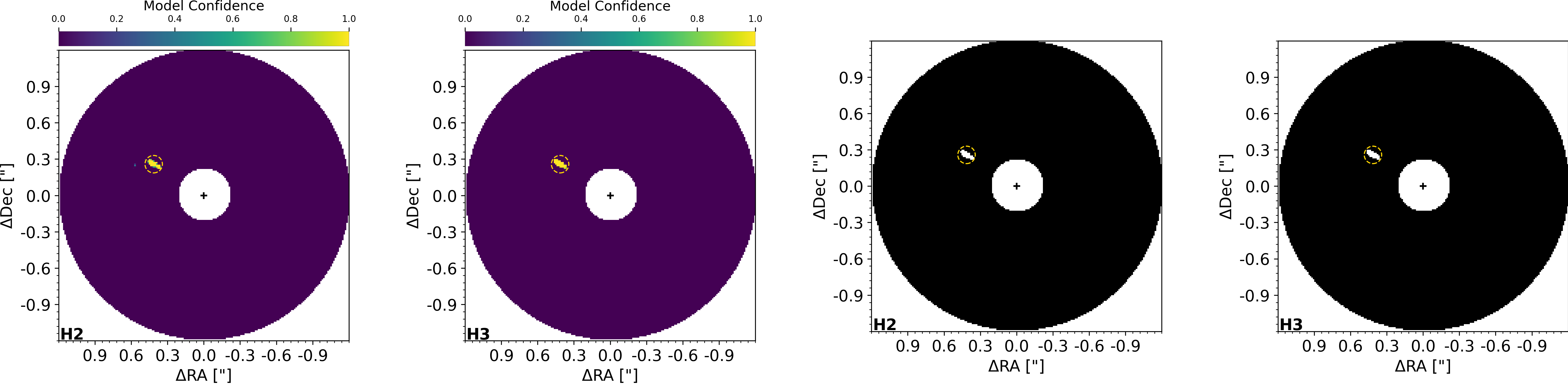} 
              \caption{Target: \textbf{PZ Tel A}\hspace{0.3cm}Observing date: \textbf{2015-05-05}\hspace{0.3cm}\label{fig:detmaps_knownComp_HIP92680}}
        \end{subfigure}\vspace{0.5cm}
        \begin{subfigure}[b]{\textwidth}
            \includegraphics[width=0.99\textwidth]{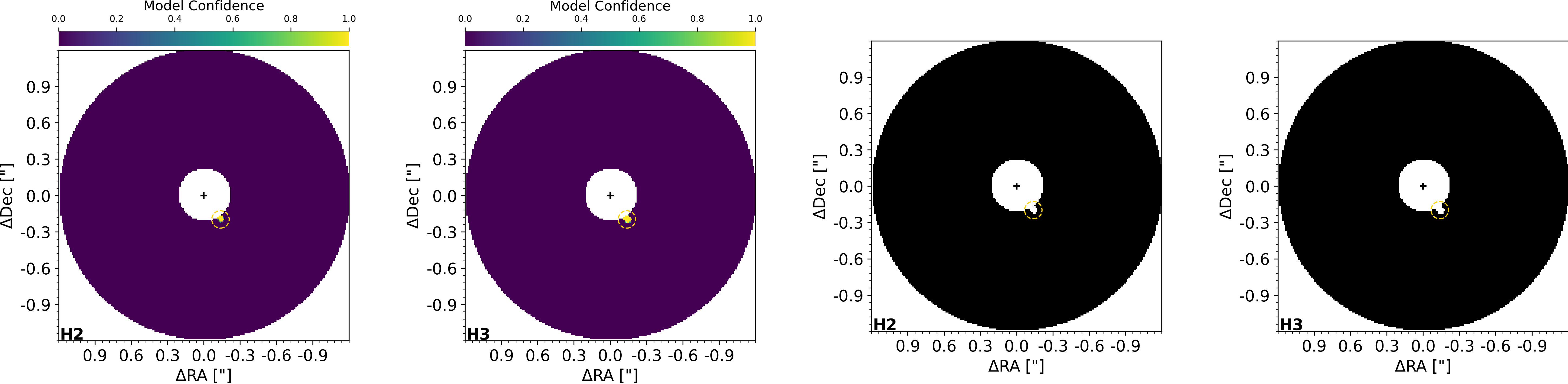} 
              \caption{Target: \textbf{$\beta$-Pictoris}\hspace{0.3cm}Observing date: \textbf{2015-12-25}\hspace{0.3cm}\label{fig:detmaps_knownComp_HIP27321}}
        \end{subfigure}
        \caption{Detection maps for a selection of F150 targets where NA-SODINN identifies confirmed substellar companions. Each horizontal panel corresponds to a different target, as indicated below the panel, and includes four sub-maps: the two on the left show NA-SODINN confidence maps for the H2 and H3 filters, while the two on the right display the corresponding \fOneScore (binary) maps, generated by applying the computed threshold mask to the confidence maps (see \cref{sec:f1_threshold}). In each sub-map, the companion is indicated with a small dashed circle in yellow, the filter used is indicated in the bottom-left corner, and the central masked region corresponds to the inner working angle. The field orientation is standard: north is up and east is to the left.}\label{fig:detection_maps_known_exoplanets}
    \end{figure*}

\begin{figure*}
        \centering
        \begin{subfigure}[b]{\textwidth}
            \includegraphics[width=0.99\textwidth]{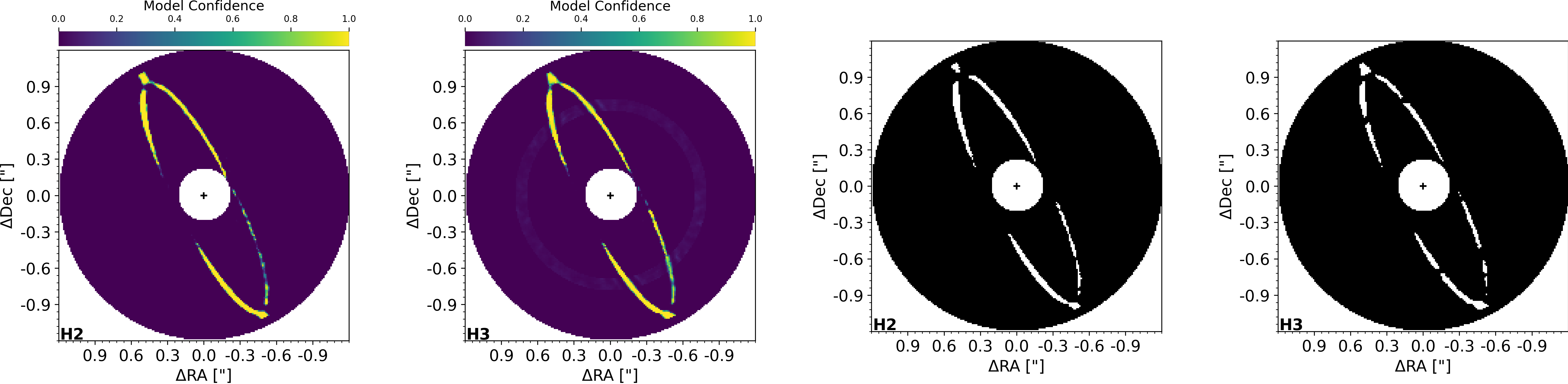}
            \caption{Target: \textbf{HD 109573}\hspace{0.3cm}Observing date: \textbf{2015-02-02}\hspace{0.3cm}\label{detmaps_knownDisks_HD109573}}
        \end{subfigure}

        \begin{subfigure}[b]{\textwidth}
            \includegraphics[width=0.99\textwidth]{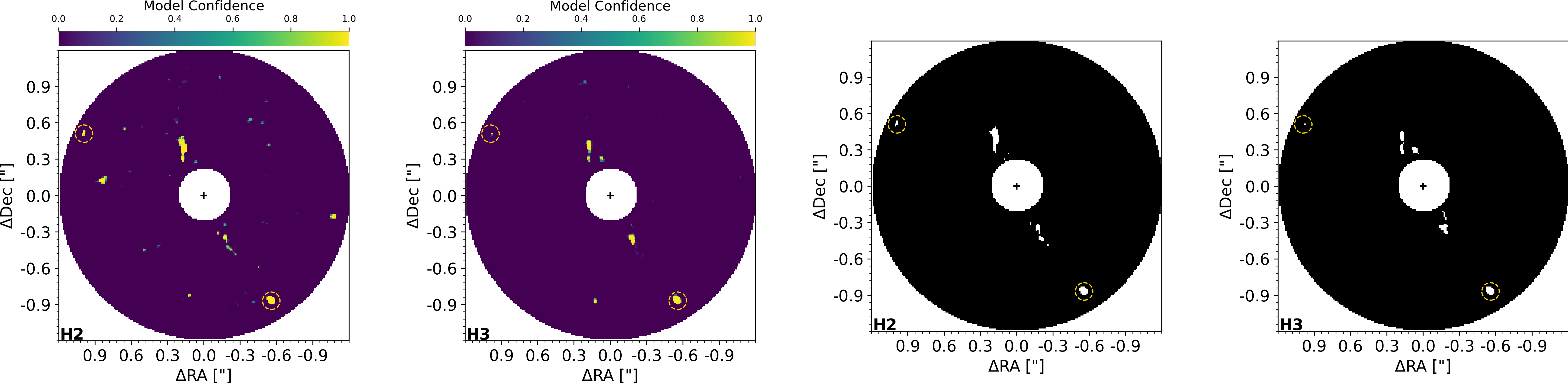}
            \caption{Target: \textbf{HD 115600}\hspace{0.3cm}Observing date: \textbf{2015-06-03}\hspace{0.3cm}\label{detmaps_knownDisks_HD115600}}
        \end{subfigure}

        \begin{subfigure}[b]{\textwidth}
            \includegraphics[width=0.99\textwidth]{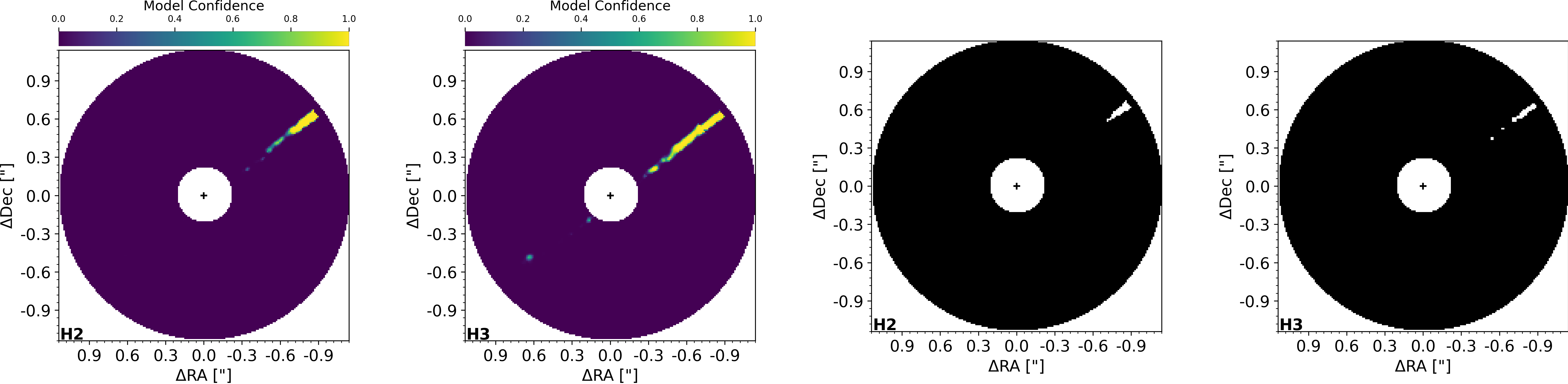}
            \caption{Target: \textbf{HD 197481}\hspace{0.3cm}Observing date: \textbf{2015-06-03}\hspace{0.3cm}\label{detmaps_knownDisks_HD197481}}
        \end{subfigure}
        
        \caption{Detection maps for the F150 sample targets in which NA-SODINN recovers known protoplanetary disks. For HD~115600 (middle panel), the two sources at large angular separations (highlighted with yellow circles), located outside the inner disk, were previously classified as background contaminants \citep{Langlois_2021_shine, Chomez_2025_shine}. The structure and interpretation of this figure follow the same format as in \cref{fig:detection_maps_known_exoplanets}.
        \label{fig:known_disks}}
    \end{figure*}

    In this appendix, we present all NA-SODINN detection maps for relevant targets in the F150 sample. The section is organized into figures, each grouping detection maps based on common features among the targets. \Cref{fig:detection_maps_known_exoplanets,fig:known_disks} present detection maps for targets with confirmed substellar companions and known protoplanetary disks, respectively. \Cref{fig:detmaps_H23_newCandidates,fig:detmaps_Honly_newCandidates} display the maps for targets where new companion candidates were identified based on the detection criteria defined in \cref{sec:f1_threshold}. Detection maps for the remaining targets are omitted, as they are largely empty and contain no detections.

    \begin{figure*}
        \centering
        \begin{subfigure}[b]{\textwidth}
            \includegraphics[width=0.99\textwidth]{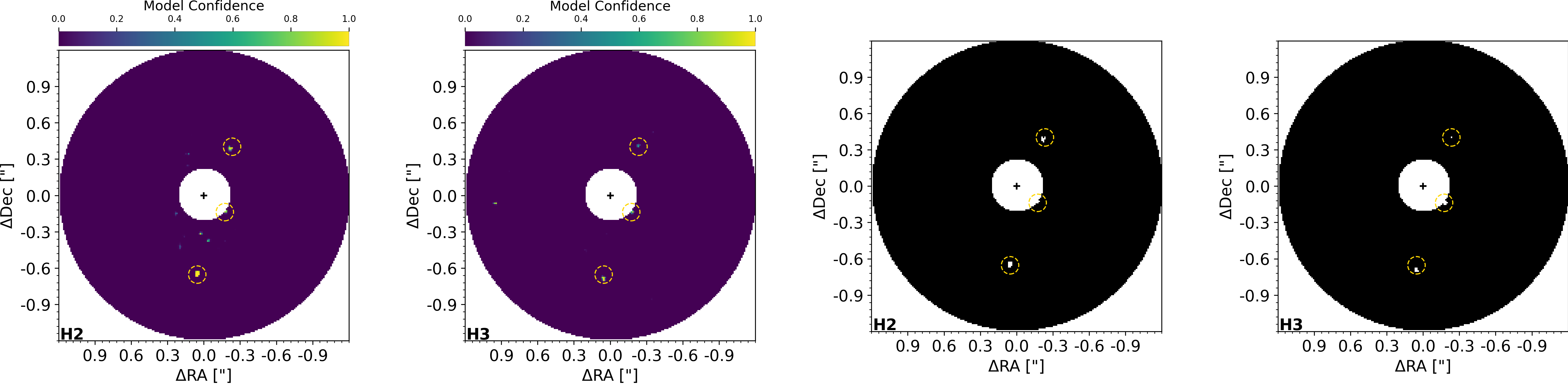} 
            \caption{Target: \textbf{Smethells 20}\hspace{0.3cm}Observing date: \textbf{2015-05-15}\hspace{0.3cm}CMD classification: \textbf{Ambiguous (c, d) and Promising (b)} \label{detmaps_target_a_2MASSJ1846}}
        \end{subfigure}\vspace{0.5cm}
        \begin{subfigure}[b]{\textwidth}
            \includegraphics[width=0.99\textwidth]{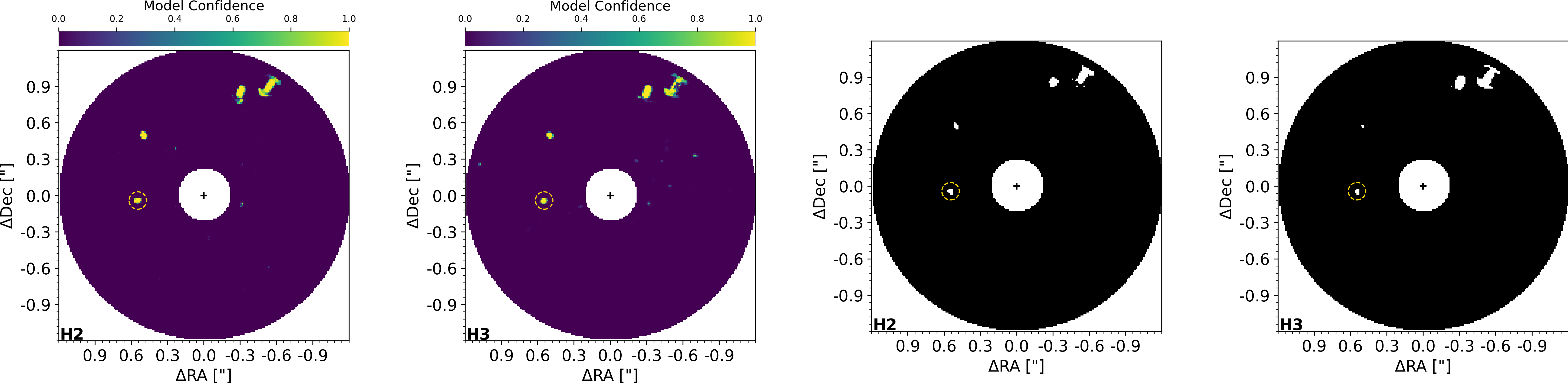} 
            \caption{Target: \textbf{HD 168210}\hspace{0.3cm}Observing date: \textbf{2017-06-02} \hspace{0.3cm} CMD classification: \textbf{Background contaminant}}
            \label{detmaps_target_b_HD168210}
        \end{subfigure}\vspace{0.5cm}
        \begin{subfigure}[b]{\textwidth}
            \includegraphics[width=0.99\textwidth]{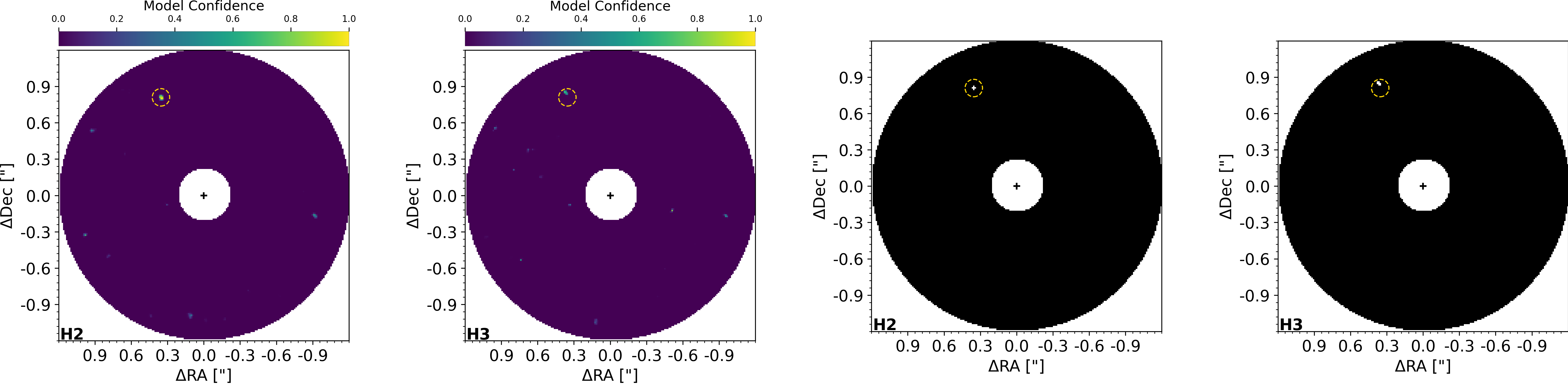} 
            \caption{Target: \textbf{$\pi$ Ara}\hspace{0.3cm}Observing date: \textbf{2016-06-10}\hspace{0.3cm} CMD classification: \textbf{Promising}}
            \label{detmaps_target_c_HIP86305}
        \end{subfigure}\vspace{0.5cm}
        \begin{subfigure}[b]{\textwidth}
            \includegraphics[width=0.99\textwidth]{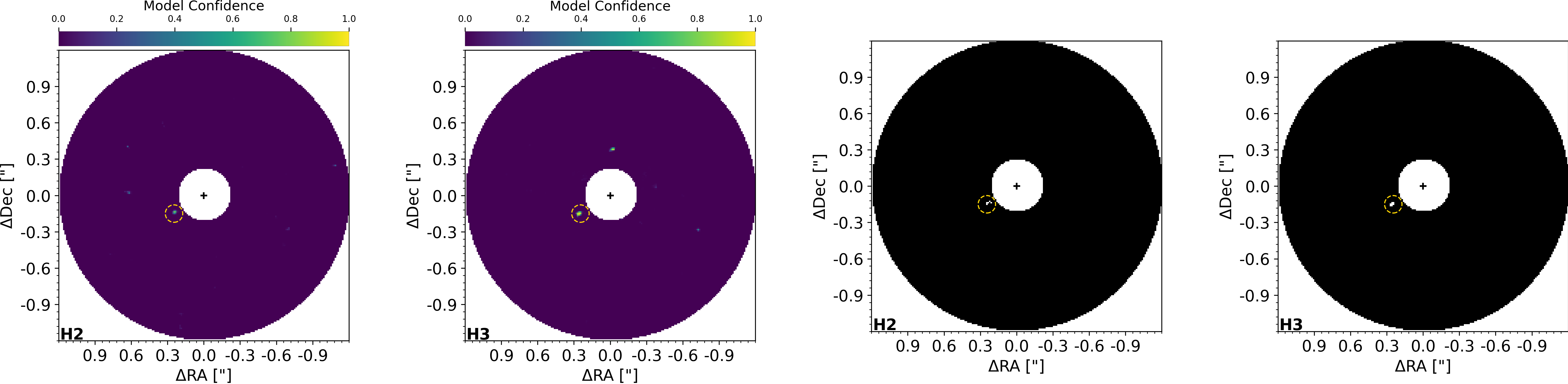} 
            \caption{Target: \textbf{CD-51 10924}\hspace{0.3cm}Observing date: \textbf{2015-05-05}\hspace{0.3cm} CMD classification: \textbf{Background contaminant}}
            \label{detmaps_target_d_HIP85647}
        \end{subfigure}
        
        \caption{Detection maps for F150 sample targets where NA-SODINN identifies new candidates, yet unreported, detected in both the H2 and H3 pass-band filters. For HD~168210 (second panel), the three additional detections at large separations (without circle) were also classified as ambiguous cases \citep{Langlois_2021_shine, Chomez_2025_shine}. Both the reference labels and the characterization of these detections can be found in \Cref{Table:targets_detections_characterization}. The structure and interpretation of this figure follow the same format as in \cref{fig:detection_maps_known_exoplanets}.}
        \label{fig:detmaps_H23_newCandidates}
    \end{figure*}

    \begin{figure*}
        \ContinuedFloat
        \centering
        \begin{subfigure}[b]{\textwidth}
            \includegraphics[width=0.99\textwidth]{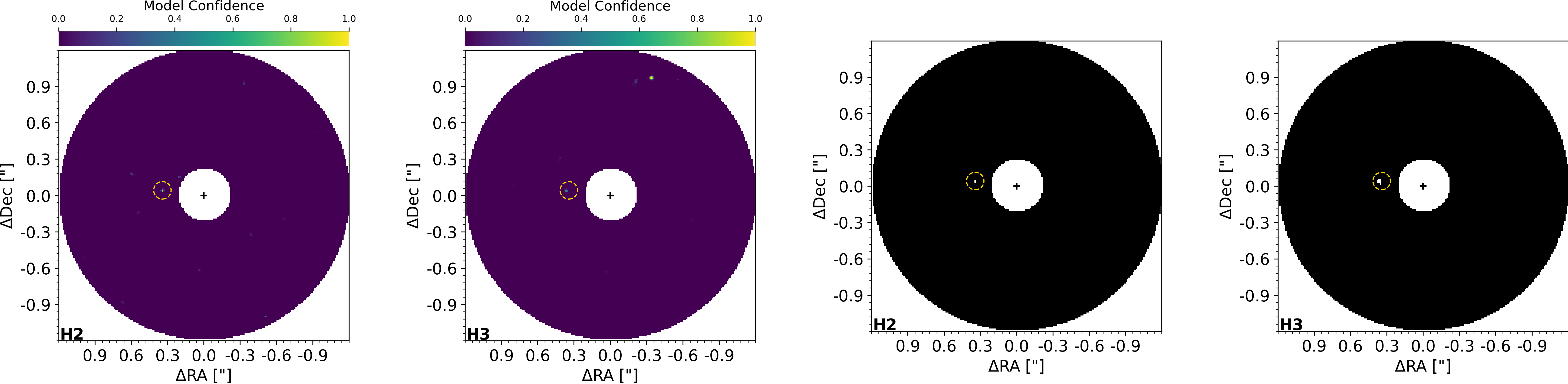} 
            \caption{Target: \textbf{$\beta$ Leo}\hspace{0.3cm}Observing date: \textbf{2015-05-30}\hspace{0.3cm} CMD classification: \textbf{Background contaminant}}
            \label{detmaps_target_e_HIP57632}
        \end{subfigure}\vspace{0.5cm}
        \begin{subfigure}[b]{\textwidth}
            \includegraphics[width=0.99\textwidth]{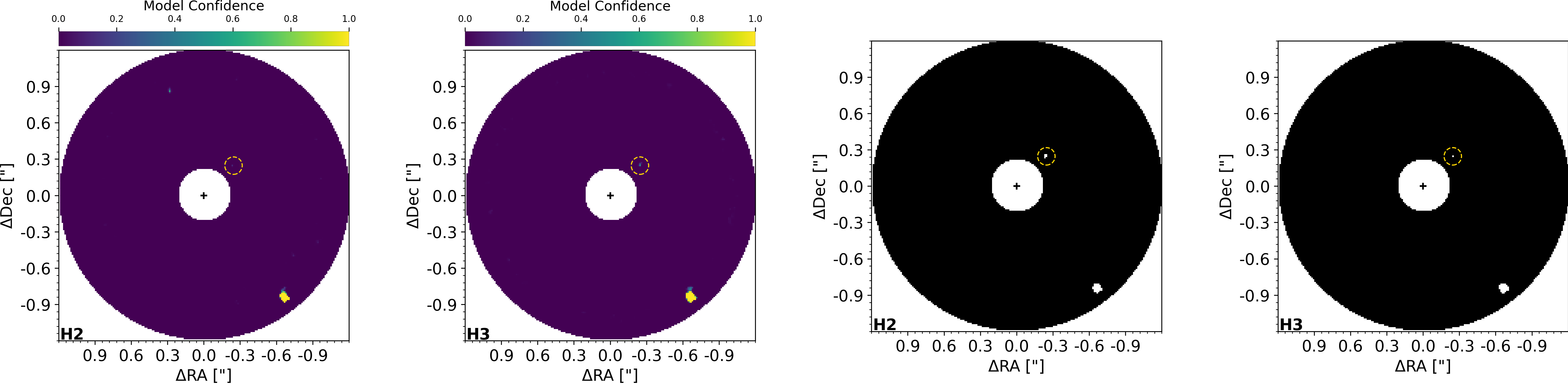} 
            \caption{Target: \textbf{HD 164249}\hspace{0.3cm}Observing date: \textbf{2018-04-10}\hspace{0.3cm} CMD classification: \textbf{Promising}}
            \label{detmaps_target_f_HD164249A}
        \end{subfigure}\vspace{0.5cm}
        \begin{subfigure}[b]{\textwidth}
            \includegraphics[width=0.99\textwidth]{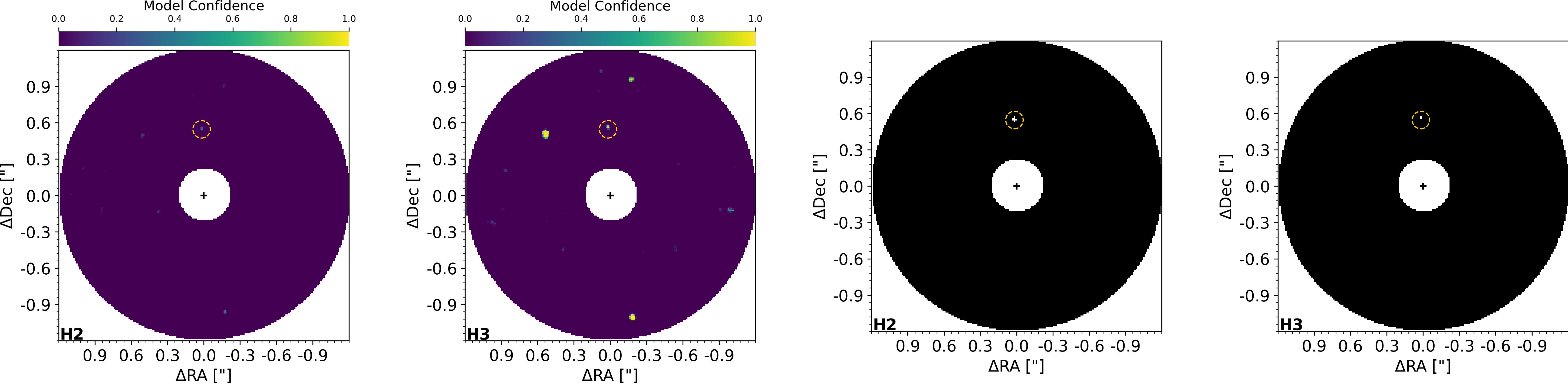} 
            \caption{Target: \textbf{CD-57 1054}\hspace{0.3cm}Observing date: \textbf{2016-01-02}\hspace{0.3cm} CMD classification: \textbf{Background contaminant}}
            \label{detmaps_target_g_HIP23309}
        \end{subfigure}
        \begin{subfigure}[b]{\textwidth}
            \includegraphics[width=0.99\textwidth]{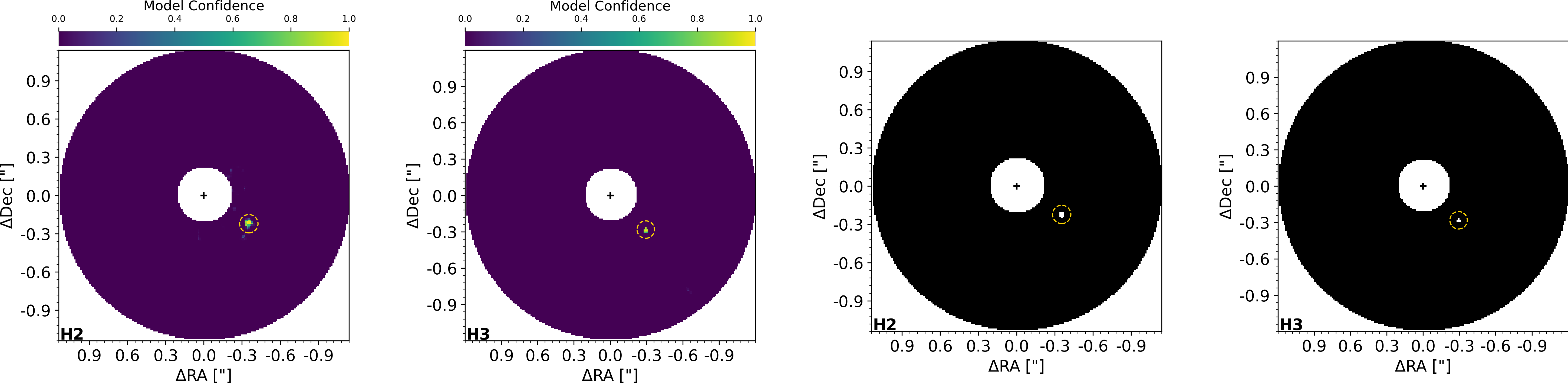} 
            \caption{Target: \textbf{CE Ant}\hspace{0.3cm}Observing date: \textbf{2017-02-06}\hspace{0.3cm} CMD classification: \textbf{Background contaminant}}
            \label{detmaps_target_g_CEAnt}
        \end{subfigure}
        
        \caption{Continuation. For HD~164249 (second panel), the candidate at large separation (without circle) was also classified as background contaminant \citep{Langlois_2021_shine, Chomez_2025_shine}.}
    \end{figure*}

    \begin{figure*}
        \begin{subfigure}[b]{\textwidth}
            \includegraphics[width=0.99\textwidth]{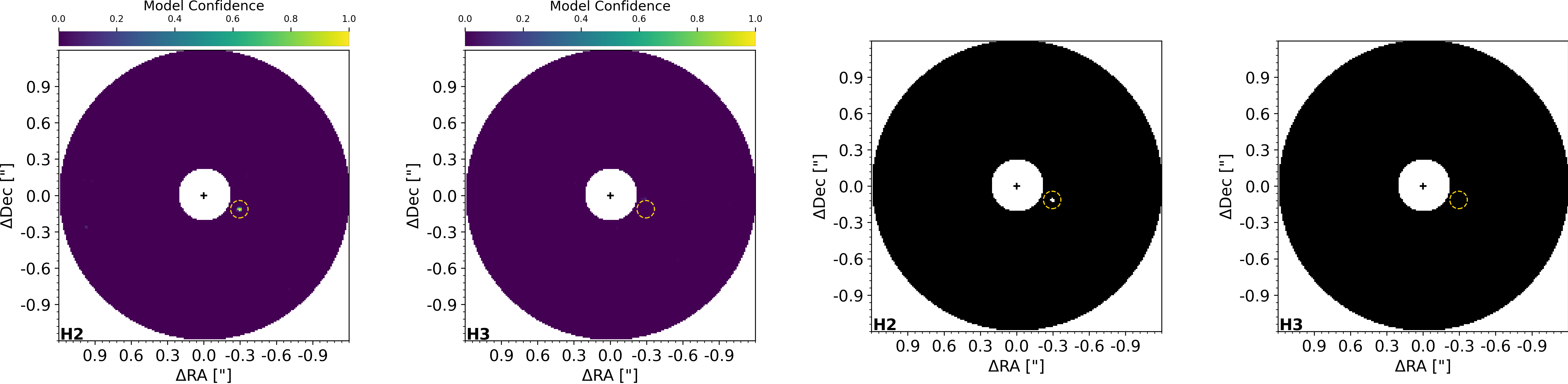} 
            \caption{Target: \textbf{CD-48 2972}\hspace{0.3cm}Observing date: \textbf{2015-11-29}\hspace{0.3cm} CMD classification: \textbf{Undefined}}
        \end{subfigure}\vspace{0.5cm}
        \begin{subfigure}[b]{\textwidth}
            \includegraphics[width=0.99\textwidth]{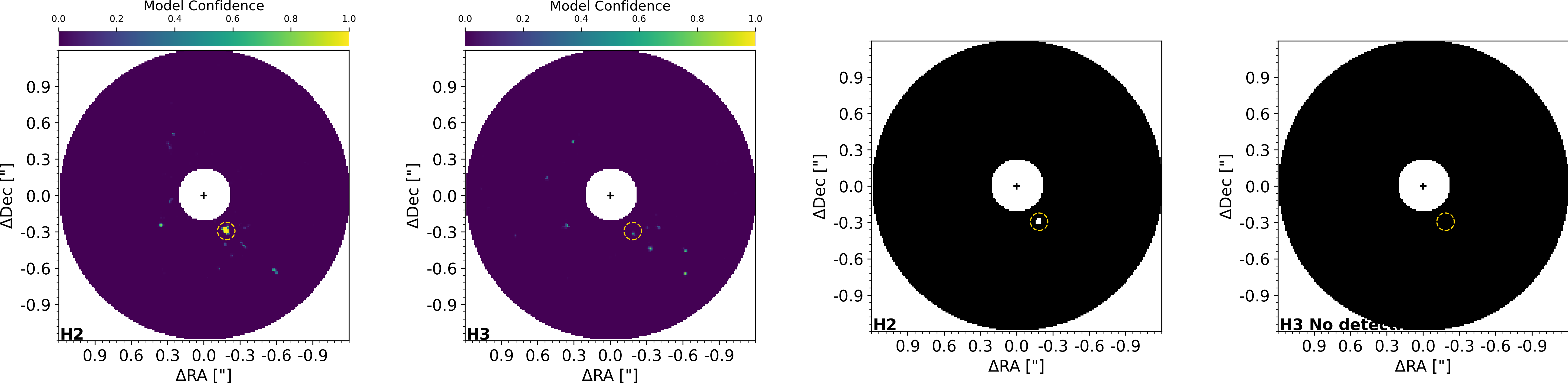} 
            \caption{Target: \textbf{HD 55279}\hspace{0.3cm}Observing date: \textbf{2017-02-07}\hspace{0.3cm} CMD classification: \textbf{Undefined}}
        \end{subfigure}\vspace{0.5cm}
        \begin{subfigure}[b]{\textwidth}
            \includegraphics[width=0.99\textwidth]{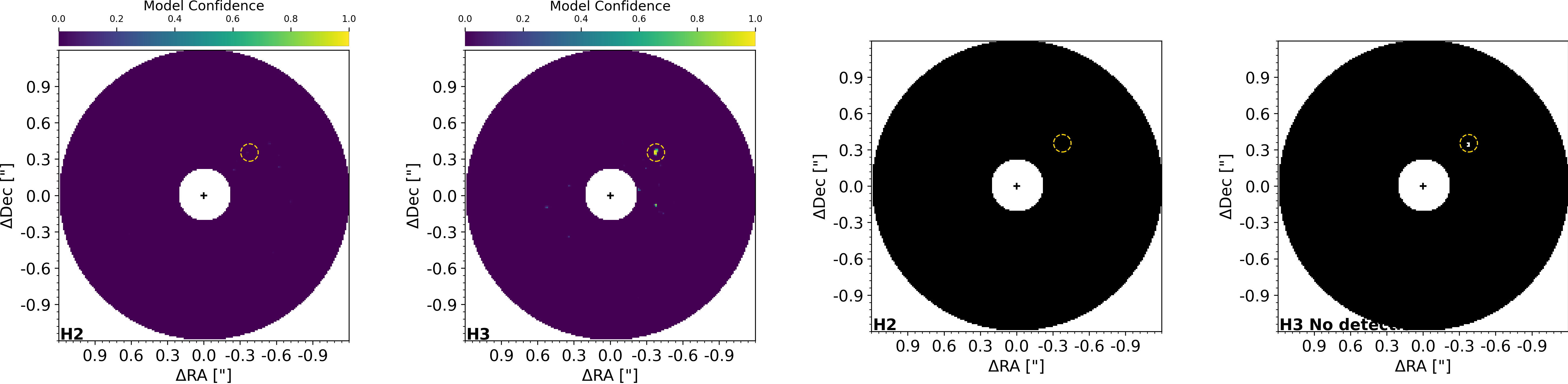} 
            \caption{Target: \textbf{HD 219482 }\hspace{0.3cm}Observing date: \textbf{2015-09-30}\hspace{0.3cm} CMD classification: \textbf{Undefined}}
        \end{subfigure}
        
        \caption{Detection maps for F150 sample targets where NA-SODINN identifies new companion candidates, detected in only one of the two passband filters (H2 or H3). The characterization of these detections can be found in \Cref{Table:targets_detections_characterization}. The structure and interpretation of this figure follow the same format as in \cref{fig:detmaps_H23_newCandidates}.}
        \label{fig:detmaps_Honly_newCandidates}
    \end{figure*}

    \section{Characterization of substellar candidates}
    \label{app:characterization_candidates}

    In this appendix, we summarize the main properties of the target stars around which NA-SODINN identified new candidates (\cref{Table:targets_detections}), together with the characterization of the candidates themselves (\cref{Table:targets_detections_characterization}).

    \clearpage

\begin{table*}[t]
    \centering
    \caption{Properties of the 11 targets around which NA-SODINN identified new companion candidates. The top panel shows targets with candidates detected in both the H2 and H3 filters, while the bottom panel shows targets with candidates detected in only one of the two filters. The two panels are ordered by decreasing right ascension (RA).}
    \label{Table:targets_detections}
    \setlength{\tabcolsep}{6pt}
    \resizebox{\textwidth}{!}{%
        \begin{tabular}{ccccccccc}
            \hline \hline
            Main Name & Other-ID & Gaia-ID & RA & DEC & H mag & Sp. type & Age [Myr] & Distance [pc] \\
            \hline
            
            CD-57 1054 & HIP 23309 & 4764027962957023104 & 05:00:47 & -57:15:25 & 6.42 & M0V & $24 \pm 5^{(1)}$ & 26.86 \\
            CE Ant & TWA 7 & 5444751795151480320 & 10:42:30 & -33:40:17 & 7.12 & M2V & $10 \pm 3^{(1)}$ & 34.10 \\
            $\beta$-Leo & HIP 57632 & - & 11:49:03 & +14:34:19 & 1.92 & A3V & $50 \pm 10^{(1)}$ & 11.00 \\
            CD-51 10924 & HIP 85647 & 5925209583053212800 & 17:30:11 & -51:38:13 & 6.08 & M0V & 15.3 & 15.98 \\
            $\pi$-Ara & HIP 86305 & 5921246427739258112 & 17:38:05 & -54:30:01 & 4.86 & A7V & $570 \pm 210^{(2)}$ & 41.01 \\
            HD 164249 & HIP 88399 & 6702775135228913280 & 18:03:03 & -51:38:56 & 6.02 & F6V & 15.3 & 49.30 \\
            HD 168210 & HIP 89829 & 4051081838710783232 & 18:19:52 & -29:16:33 & 7.19 & G5V & $24 \pm 5^{(1)}$ & 80.46 \\
            Smethells 20 & TYC 9073-762-1 & 6631685008336771072 & 18:46:52 & -62:10:36 & 8.04 & M1V & $24 \pm 5^{(1)}$ & 50.69 \\
            
            \hline
            HD 55279 & HIP 33737 & 5208216951043609216 & 07:00:30 & -79:41:45 & 7.83 & K2V & $45 \pm 10^{(1)}$ & 63.45 \\
            CD-48 2972 & TYC 8128-1946-1 & 5506101790904438656 & 07:28:22 & -49:08:37 & 8.13 & G8V & $50 \pm 10^{(1)}$ & 87.21 \\
            HD 219482 & HIP 114948 & 6489909443564308224 & 23:16:57 & -62:00:04 & 4.60 & F6V & $300 \pm 150^{(2)}$ & 20.44 \\
            \hline
        \end{tabular}%
    }

    \vspace{0.3cm}
    \begin{minipage}{0.95\textwidth}
    \footnotesize
    \textbf{References.} Ages are adopted from different sources:
    (1) based on membership in moving groups, associations, or clusters;
    (2) from \citet{Desidera_2021_shine}.
    Distances are derived from \textit{Gaia} parallaxes.
    \end{minipage}
\end{table*}

    \begin{table*}[t]
        \centering
        \caption{Characterization of the newly identified candidates detected by NA-SODINN, as shown in \cref{fig:detmaps_H23_newCandidates,fig:detmaps_Honly_newCandidates}. The table is divided into eleven sections, each corresponding to a target listed in \cref{Table:targets_detections}, around which one or more companions are identified. The column CMD reports the candidate classification based on the analysis in \cref{sec:new_candidates}.}
        \label{Table:targets_detections_characterization}
        
        \setlength{\tabcolsep}{8pt}

        \begin{tabular}{ccccccc}
        \hline\hline
        Candidate ID & Band & Separation [mas] & Position angle [$^\circ$] & Contrast & $\Delta$mag & CMD \\
        \hline
        
        \multicolumn{7}{c}{\textbf{Smethells 20}} \\
        b & H2 & $218.94 \pm 30.02$ & $323.35 \pm 7.31$ & $1.04 \pm 0.55 \times 10^{-4}$ & $9.95 \pm 0.65$ & \textbf{Promising} \\
          & H3 & $227.59 \pm 29.21$ & $322.37 \pm 7.09$ & $7.65 \pm 3.87 \times 10^{-5}$ & $10.29 \pm 0.69$ & \\
        c & H2 & $448.36 \pm 30.64$ & $60.65 \pm 3.51$ & $2.68 \pm 1.43 \times 10^{-5}$ & $11.33 \pm 0.79$ & Ambiguous \\
          & H3 & $462.80 \pm 30.05$ & $60.46 \pm 3.27$ & $2.39 \pm 1.53 \times 10^{-5}$ & $11.42 \pm 0.74$ & \\
        d & H2 & $664.86 \pm 28.94$ & $264.69 \pm 2.35$ & $2.55 \pm 1.59 \times 10^{-5}$ & $11.34 \pm 0.72$ & Ambiguous \\
          & H3 & $700.81 \pm 29.74$ & $264.32 \pm 2.29$ & $1.83 \pm 1.03 \times 10^{-5}$ & $11.91 \pm 0.83$ & \\
        
        \multicolumn{7}{c}{\textbf{HD 168210}} \\
        b & H2 & $569.35 \pm 21.03$ & $185.00 \pm 1.10$ & $3.82 \pm 2.17 \times 10^{-6}$ & $13.55 \pm 0.61$ & Background \\
          & H3 & $568.91 \pm 26.80$ & $185.15 \pm 2.55$ & $5.33 \pm 2.42 \times 10^{-6}$ & $13.19 \pm 0.50$ & \\
        
        \multicolumn{7}{c}{\textbf{$\pi$ Ara}} \\
        b & H2 & $888.45 \pm 21.75$ & $114.26 \pm 1.16$ & $6.31 \pm 2.85 \times 10^{-6}$ & $13.00 \pm 0.51$ & \textbf{Promising} \\
          & H3 & $920.71 \pm 25.46$ & $114.36 \pm 1.28$ & $4.67 \pm 3.24 \times 10^{-6}$ & $13.33 \pm 0.79$ & \\
        
        \multicolumn{7}{c}{\textbf{CD-51 10924}} \\
        b & H2 & $280.09 \pm 27.01$ & $210.62 \pm 5.14$ & $2.22 \pm 1.43 \times 10^{-5}$ & $11.63 \pm 0.74$ & Background \\
          & H3 & $294.78 \pm 26.33$ & $210.21 \pm 4.27$ & $1.46 \pm 0.91 \times 10^{-5}$ & $12.08 \pm 0.76$ & \\
        
        \multicolumn{7}{c}{\textbf{$\beta$ Leo}} \\
        b & H2 & $370.09 \pm 23.98$ & $175.30 \pm 2.87$ & $1.48 \pm 0.62 \times 10^{-5}$ & $12.07 \pm 0.50$ & Background \\
          & H3 & $377.84 \pm 24.67$ & $178.50 \pm 4.15$ & $2.68 \pm 1.15 \times 10^{-5}$ & $11.43 \pm 0.47$ & \\
        
        \multicolumn{7}{c}{\textbf{HD 164249}} \\
        b & H2 & $330.35 \pm 30.77$ & $47.21 \pm 5.09$ & $2.66 \pm 1.13 \times 10^{-5}$ & $11.43 \pm 0.74$ & \textbf{Promising} \\
          & H3 & $341.74 \pm 31.05$ & $46.26 \pm 4.58$ & $1.97 \pm 0.87 \times 10^{-5}$ & $11.34 \pm 0.79$ & \\
        
        \multicolumn{7}{c}{\textbf{CD-57 1054}} \\
        b & H2 & $552.91 \pm 30.26$ & $92.61 \pm 2.74$ & $7.60 \pm 3.78 \times 10^{-6}$ & $12.80 \pm 0.52$ & Background \\
          & H3 & $560.06 \pm 29.50$ & $92.32 \pm 2.51$ & $6.95 \pm 4.12 \times 10^{-6}$ & $12.90 \pm 0.68$ & \\
        
        \multicolumn{7}{c}{\textbf{CE Ant}} \\
        b & H2 & $413.34 \pm 28.64$ & $328.7 \pm 3.79$ & $4.20 \pm 1.97 \times 10^{-6}$ & $13.43 \pm 0.62$ & Background \\
          & H3 & $423.69 \pm 28.97$ & $328.36 \pm 3.68$ & $3.90 \pm 1.80 \times 10^{-6}$ & $13.53 \pm 0.51$ & \\
        
        \multicolumn{7}{c}{\textbf{CD-48 2972}} \\
        b & H2 & $305.32 \pm 31.39$ & $337.29 \pm 5.55$ & $1.73 \pm 1.01 \times 10^{-5}$ & $11.90 \pm 0.63$ & Undefined \\
        
        \multicolumn{7}{c}{\textbf{HD 55279}} \\
        b & H2 & $318.60 \pm 27.84$ & $302.87 \pm 4.23$ & $2.91 \pm 1.57 \times 10^{-5}$ & $11.34 \pm 0.62$ & Undefined \\
        
        \multicolumn{7}{c}{\textbf{HD 219482}} \\
        b & H3 & $399.89 \pm 24.64$ & $47.17 \pm 3.22$ & $2.00 \pm 1.23 \times 10^{-6}$ & $14.23 \pm 0.75$ & Undefined \\
        
        \hline
        \end{tabular}
    \end{table*}

\end{appendix}
    
\end{document}